\documentclass[%
reprint,
superscriptaddress,
nofootinbib,
amsmath,amssymb,aps,prx]{revtex4-2}

\usepackage{graphicx}% Include figure files
\usepackage{dcolumn}% Align table columns on decimal point
\usepackage{bm}
\usepackage{tikz}
\usepackage{dsfont}
\usepackage[caption=false]{subfig}
\usepackage{physics}
\usetikzlibrary{calc,positioning,decorations.pathreplacing,calligraphy,matrix,graphs,fit,shapes,shapes.geometric,arrows.meta,patterns}
\tikzset{sss/.tip={[sep=1pt 1]
		Butt Cap[] . Square[length=0pt 1] Square[length=0pt 1] Square[length=0pt 1, sep=0pt]}}
\tikzset{>=latex}

\newcommand{\rme}{\mathrm{e}}
\newcommand{\inta}[1]{\int_0^{\infty}\mathrm{d}#1\ }

\newcommand{\mb}{\mathbf}

\newcommand{\sket}[1]{\ket{#1}\rangle}

\newcommand{\timo}{\overleftarrow{T}}

\newcommand{\uk}{_{\mathbf{k}}}

\newcommand{\uq}{_{\mathbf{q}}}
\newcommand{\uqp}{_{\mathbf{q}^{\prime}}}

\usepackage{etoolbox}
\preto\subequations{\ifhmode\unskip\fi} %prevents white space before \subequations

\begin{document}
	
	\preprint{APS/123-QED}
	
	\title{Multiple-bath extension of TEMPO}
	\title{Exact dynamics of non-additive environments in non-Markovian open quantum systems}
	\author{Dominic Gribben}
		\thanks{These authors contributed equally.}
	\affiliation{SUPA, School of Physics and Astronomy, University of St Andrews, St Andrews, KY16 9SS, UK}

	\author{Dominic M. Rouse}
		\thanks{These authors contributed equally.}
	\affiliation{SUPA, School of Physics and Astronomy, University of St Andrews, St Andrews, KY16 9SS, UK}
	
	\author{Jake Iles-Smith}
		\thanks{These authors contributed equally.}
	\affiliation{Department of Physics and Astronomy, The University of Manchester,
Oxford Road, Manchester, M13 9PL, UK}
\affiliation{
Department of Electrical and Electronic Engineering, The University of Manchester,
Sackville Street Building, Manchester, M1 3BB, UK}

	\author{Aidan Strathearn}
	\affiliation{School of Mathematics and Physics, The University of Queensland, St Lucia,
Queensland 4072, Australia}
	
	\author{Henry Maguire}
	\affiliation{Department of Physics and Astronomy, The University of Manchester,
Oxford Road, Manchester, M13 9PL, UK}
	
	\author{Peter Kirton}
	\affiliation{Department of Physics and SUPA, University of Strathclyde, Glasgow G4 0NG, United Kingdom}
	
	\author{Ahsan Nazir}
	\affiliation{Department of Physics and Astronomy, The University of Manchester,
Oxford Road, Manchester, M13 9PL, UK}
	
	\author{Erik M. Gauger}
	\affiliation{SUPA, Institute of Photonics and Quantum Sciences, Heriot-Watt University, Edinburgh, EH14 4AS, UK}
	
	\author{Brendon W. Lovett}
	\email{bwl4@st-andrews.ac.uk}
	\affiliation{SUPA, School of Physics and Astronomy, University of St Andrews, St Andrews, KY16 9SS, UK}

	\date{\today}% 
	
	\begin{abstract}
        When a quantum system couples strongly to multiple baths then it is generally no longer possible to describe the resulting system dynamics by simply adding the individual effects of each bath. However, capturing such multi-bath system dynamics has up to now required approximations that can obscure some of the non-additive effects. Here we present a numerically-exact and efficient technique for tackling this problem that builds on the time-evolving matrix product operator (TEMPO) representation. We test the method by applying it to a simple model system that exhibits non-additive behaviour: a two-level dipole coupled to both a vibrational and an optical bath. Although not directly coupled, there is an effective interaction between the baths mediated by the system that can lead to population inversion in the matter system when the vibrational coupling is strong. 
        We benchmark and validate multi-bath TEMPO against two approximate methods --- one based on a polaron transformation, the other on an identification of a reaction coordinate --- before exploring the regime of simultaneously strong vibrational and optical coupling where the approximate techniques break down. Here we uncover a new regime where the quantum Zeno effect leads to a fully mixed state of the electronic system.
	
		%Recently, a numerically exact and efficient technique using time-evolving matrix product operators (TEMPO) was developed to describe the time-evolution of a quantum system coupled to a bosonic bath. Here, we extend this technique to allow coupling to multiple non-commuting baths. To demonstrate the efficacy of this, we explore the physics of a two level system coupled to both an optical and vibrational bath. In this system, the effects of the baths are non-additive and can lead to population inversion in the matter system when the vibrational coupling is strong enough. To compare to TEMPO, we use the reaction coordinate and polaron transformation techniques to probe analytically the origin of the population inversion in the strong vibrational and weak optical coupling regime. We find that the non-additivity leads to renormalisation of the optical transition rates, which ultimately stems from the overlap of the displaced vibrational states as well as the energy separations. Whether or not population inversion can occur is strongly dependent on the structure of the optical spectral density. Finally, using TEMPO we find that the population inversion is suppressed as the optical coupling is also increased.
	\end{abstract}
	
	\maketitle
	
	\section{Introduction}
	
	Open quantum systems are often significantly coupled to more than one kind of environment. However, the combined influence of these different environments is generally more than the sum of their individual parts: it is thus crucial to account for \emph{non-additive} effects --- that is, effects that originate from an interplay between two or more competing environments~\cite{Giusteri2017inter, PhysRevA.97.062124,mitchison2018non,maguire2019environmental,McConnell2019count}.

	Examples of non-additive behaviour include those seen or predicted in optically active quantum systems that are strongly influenced by their vibrational environments, which are ubiquitous in condensed matter and molecular physics. 
	Not only do vibrational interactions lead to complex dynamical behaviour, they are central in determining the optical and electronic properties of a system. For example, they are thought to play a key role in light harvesting and energy transfer   
	%Optically active molecules with vibrational environments are a commonly studied system due to their key roles in light harvesting and energy transfer 
	\cite{hedley2017light,adronov2000light, felip2016chameleonic, burzuri2016sequential,scholes2011lessons}. Another example occurs in molecular nanojunctions, where the combined effect of the leads and vibrational environments is non-additive~\cite{thomas2019understanding, sowa2020beyond}, and using an additive treatment can even lead to a violation of the Carnot bound on efficiency~\cite{mcconnell2021strong}. Using additive treatments can lead to other unphysical predictions too, for example anomalous emission of photons from the groundstate in regimes of strong light-matter coupling~\cite{Ciuti06}. Further, it can miss key dynamical and steady state behaviour, such as the Franck-Condon blockade observed in quantum transport~\cite{Koch2005Frank,Koch2006blockade}.

%	Not only do vibrational interactions lead to complex dynamical behaviour, they are central in determining the optical and electronic properties of a system. For example, they are thought to play a key role in light harvesting and energy transfer   
	%Optically active molecules with vibrational environments are a commonly studied system due to their key roles in light harvesting and energy transfer 
	%\cite{hedley2017light,adronov2000light, felip2016chameleonic, burzuri2016sequential,scholes2011lessons}. 
	
	\begin{figure*}[t]
	\includegraphics[width=\textwidth]{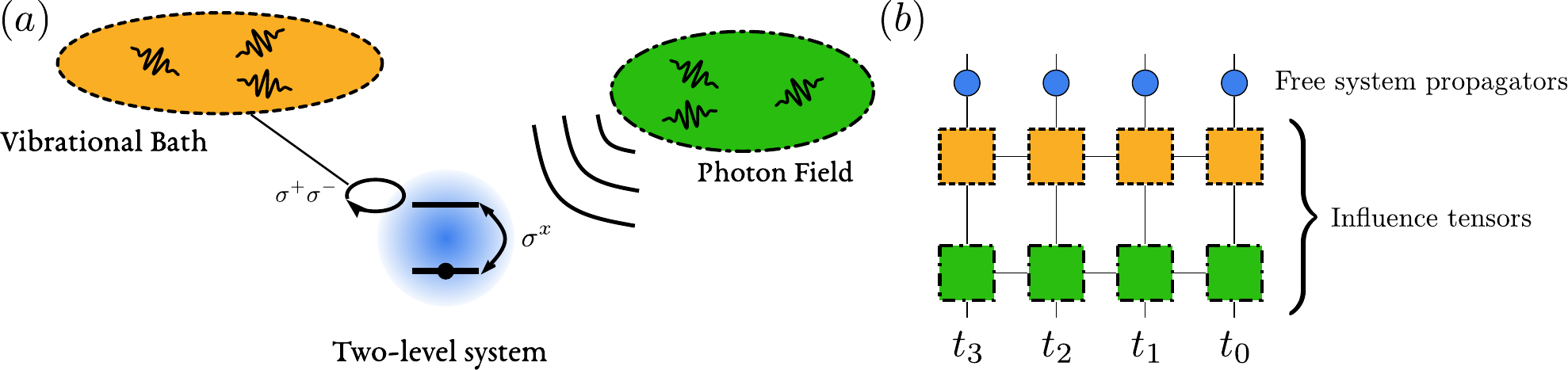}
	\caption{\label{fig:cartoon}(a) Cartoon of the model considered in this paper: a two-level dimer coupled to both a vibrational bath and a photon field. (b) schematic tensor network representation of our technique for non-perturbative simulation of the dynamics. By considering how the system evolves timestep-by-timestep there is a contribution from both the time-local free system propagators and the influence tensors for each environment which account for any non-Markovian effects induced by the interactions.}
	\end{figure*}
	
	However, typical standard perturbative methods such as Redfield theory do make an implicit additive approximation, such that the impact of each individual environment
	% (in this case, optical and vibrational) 
	is mutually independent.
	Such a treatment will fail in general once the magnitude of the system-bath interaction Hamiltonians for each environment become strong, when non-Markovian effects are important.
%	This can lead to unphysical predictions, for example anomalous emission of photons from the groundstate in regimes of strong light-matter coupling~\cite{Ciuti06}, or it can miss key dynamical and steady state behaviour, such as the Frank-Condon blockade observed in quantum transport~\cite{Koch2005Frank,Koch2006blockade}.
 A number of techniques have been developed to capture non-additive and strong coupling effects, while maintaining the conceptual simplicity of Born-Markov master equations, such as the polaron transformation \cite{mccutcheon2010quantum,roy2011phonon,roy2012polaron,wilson2002quantum,qin2017effects,rouse2019optimal} and its extensions \cite{pollock2013multi, mccutcheon2011variational,bundgaardnielsen2021nonmarkovian}; methods based on the pseudomode approach~\cite{Garraway1997pseudo,Dalton2001pseudo,Mazzola2009pseudo} and its generalisation~\cite{Tamascelli2018nonpert,Pleasance2020pseudo,Mascherpa2020aux}; and the reaction coordinate (RC) mapping \cite{maguire2019environmental,iles2016energy,garg1985effect}. 
	Even though these methods can access non-perturbative regimes, they are typically limited to a particular parameter regime due to the approximations required in their derivation.
	%An essential ingredient of these techniques is that they capture the non-additive nature of the optical and vibrational interactions, an effect that becomes increasingly prominent at strong coupling and is completely overlooked by standard perturbative approaches. 
	
    To overcome these kinds of limitations, tensor network techniques have become increasingly useful in reducing the computational cost and time of numerically exact calculations of open quantum system dynamics~\cite{cygorek2021numericallyexact}. 
    Although originally formulated to more efficiently represent many-body states with area law entanglement~\cite{orus2014practical,orus2019tensor} they are of more general use in problems which can be cast in a form where the compression of large tensors gives a significant numerical advantage. For example, chain-mapping techniques~\cite{chin2010exact,prior2010efficient} involve transforming the topology of system-bath couplings from a star to a 1D chain with nearest-neighbour couplings; this can then be efficiently represented as a spatial matrix product state (MPS). 
    An alternative approach is to construct an MPS in time able to efficiently capture non-Markovian correlations in an equivalent fashion that spatial MPS capture spatial correlations~\cite{strathearn2017efficient,pollock2018nonmarkovian}. 
    The resulting time-evolving matrix product operator (TEMPO) method provides an efficient approach for calculating the exact dynamics of an open quantum system interacting with a complex environment.
    
    In this work we describe how to formulate {\it multi-bath} TEMPO, and so describe a numerically exact approach for capturing the effects of strongly coupling environments that are non-additive.
    As an example calculation using this new approach, we
    investigate the nonequilibrium behaviour of a two-level quantum emitter interacting with both vibrational and optical environments.
    We find that multi-bath TEMPO allows us to explore parameter regimes inaccessible to approximate techniques.
    We first confirm the behaviour predicted in Ref.~\cite{maguire2019environmental}, where non-additive effects lead to a population inversion if the optical temperature and vibrational coupling strength are large enough. 
    %We further show that the ability of a system to exhibit population inversion is determined by the optical spectral density, which is itself dependent on the gauge choice \cite{stokes2012extending,stokes2018master}. 
    %We find that the population inversion is suppressed at sufficiently large optical coupling. 
    %We suggest that the mechanism for this is a \emph{dynamical undressing} of phonons from the electronic states, which occurs when optical timescales are short enough to prevent polaron formation. This effect is equivalent to that seen in both classically driven systems~\cite{nazir2016modelling,strathearn2020modelling} and systems coupled to a finite number of cavity modes~\cite{denning2020optical} [Reiter?].
We then test multi-bath TEMPO in the previously inaccessible regime of simultaneously strong optical and vibrational coupling. Here we find the steady state population of both states of the emitter is fixed at one half regardless of temperature; we associate this with a quantum Zeno effect arising from the strong optical coupling. 
    This serves to not only prove the importance of non-additive effects in the behaviour of open quantum systems, but also reveals new unexpected behaviour in the nonequilibrium steady state of the system. By comparing TEMPO to the polaron and RC theories, we gain analytic insight into the otherwise purely numerical approach, as well as determine when such approximate techniques are accurate.
    % \jake{Where do Gauges fit in all of this? Perhaps the approximate methods are taking a back seat here?} \dom{I've added a bit about gauges in this paragraph. Also, added a one liner at the end here to tie back together approximate methods and TEMPO. But it might be okay if approximate methods take a bake seat since we are focusing on tempo now?}
    
   In the next section we will describe the multi-bath TEMPO technique in detail. In Section~\ref{Sec:modelandmethods} we will set out the example calculation, introducing the model of a single dipole coupled to both an optical and vibrational bath. We will also show, in a simplified description, the mechanism by which population inversion can occur. In Section~\ref{Sec:approxtech}, we summarise the polaron transformation and RC mapping techniques, which we will compare to the exact results obtained from multiple-bath TEMPO in Section~\ref{Sec:Comparisons}. 
   %The comparison we make is of the steady state of the 
   We explore when the approximate techniques are accurate, and present results that can only be captured using our new approach. Our conclusions are drawn in Section~\ref{Sec:conc}.

	\section{Multiple-bath TEMPO}\label{sec:TEMPO}
	In this section we present a derivation of the TEMPO algorithm extended to simulate a system coupled to multiple baths with non-commuting operators. A key step in the original path integral framework~\cite{makri1995tensor,makri1995tensor-2,strathearn2017efficient} is to insert the complete eigenbasis of the coupling operator between each timestep of a discretized influence functional.
	 A different approach must be taken in the general case of multiple coupling operators that do not share an eigenbasis, and this is discussed in Ref.~\cite{strathearn2020modelling}. However, a viable method of contracting the resulting influence tensor is still required, and we provide this here.
	In particular, we enable efficient contraction of this influence tensor by decomposing it into contributions from each environment, allowing us to construct the influence tensor for each independently. This is distinct from the independent calculation, and subsequent addition, of dynamical generators. In combining the separate influence tensors the total, non-additive effect of the environments is faithfully recovered provided this is done on a fine enough timescale. In Fig.~\ref{fig:cartoon} we show this process schematically for the specific model we will consider from Sec.~\ref{Sec:model} of this paper: a two level system coupled to two environments. 
	%In this paper we only wish to make predictions about the system steady state behaviour, which we will extract by calculating dynamics for times long enough for transients to die away. We therefore only contract what is necessary to build the required dynamical maps.
	This process is made tractable by employing efficient tensor network representations~\cite{orus2014practical,orus2019tensor} of both the separate environmental contributions and the combined effect of all these environments and the internal system Hamiltonian ~\cite{pollock2018nonmarkovian,jorgensen2019exploiting}.

	\subsection{Multi-bath influence functional}
	The generic Hamiltonian we consider is
	\begin{equation}\label{eq:genham}
	H=H_S+\sum_{\alpha}(H_{I\alpha}+H_{B\alpha}),
	\end{equation} 
	where $H_S$ is the (arbitrary) system Hamiltonian, and
	\begin{equation}
	H_{I\alpha}= s_{\alpha}\sum_q\left(g_{\alpha q}  a_{\alpha q} + g_{\alpha q}^* a_{\alpha q}^\dagger\right),
	\end{equation}
	and
	\begin{equation}
 H_{B\alpha}= \sum_q\omega_{\alpha q} a_{\alpha q}^\dagger a_{\alpha q},
	\end{equation}
	are the interaction and free bath Hamiltonians, respectively. Here $a_{\alpha q}$ ($a_{\alpha q}^\dagger$) is the annihilation (creation) operator corresponding to mode $q$ in bath $\alpha$ with frequency $\omega_{\alpha q}$. The arbitrary system operator $s_{\alpha}$ couples to mode $q$ in bath $\alpha$ with strength $g_{\alpha q}$. The baths can be continuous or discrete; in this paper we consider two continua. We assume that the initial state is separable into system and bath terms with the baths each initially in a Gaussian state e.g.~thermal equilibrium at temperatures $T_{\alpha}$. 
	
	We work in a representation where the $d\times d$ density matrix $\rho$, an operator on the Hilbert space $\mathcal{H}$ spanned by basis vectors $\{\ket{e_i}\}$, is mapped to a $d^2$ vector in Liouville space $\mathfrak{L}=\mathcal{H}\otimes \mathcal{H}$ spanned by $\{\ket{e_i}\otimes\ket{e_j}\}$. Formally this mapping corresponds to the following:
	\begin{align}
	    \rho=\sum_{ij}\rho_{ij} \dyad{e_i}{e_j} &\rightarrow \sket{\rho}=\sum_{ij} \rho_{ij} \ket{e_i}\otimes\ket{e_j}, \\
	    O\rho &\rightarrow O^L \sket{\rho}=O\otimes\mathbb{I}\sket{\rho},\\
	    \rho O &\rightarrow O^R \sket{\rho}=\mathbb{I}\otimes O^T \sket{\rho}.
	\end{align}
	For convenience we proceed in the interaction picture with respect to $H_S$ and $H_{B\alpha}$. The von Neumann equation is then
	\begin{equation}
	   \frac{d}{dt}\sket{\rho(t)}=\sum_{\alpha}\mathcal{L}_{I\alpha}(t)\sket{\rho(t)},
	\end{equation}
	where we have introduced the Liouvillian superoperators $\mathcal{L}_{I\alpha}  = -i( H_{I\alpha}^L -  H_{I\alpha}^{R})$. The exact solution of the von Neumann equation can be written as
	\begin{align}\label{eq:vnsol}
	\sket{\rho (t)} &= \timo \exp \left(\int_0^t \sum_{\alpha} \mathcal{L}_{I\alpha} (t') dt'\right) \sket{\rho(0)} \nonumber \\ &= \Lambda(t)\sket{\rho(0)},
	\end{align}
	where $\timo$ signifies that superoperators be time-ordered from right to left and $\Lambda(t)$ is the propagator. We now factorise the propagator into a product of exponentials for each bath,
	\begin{equation}\label{eq:splitprop}
	\Lambda(t) = \timo \prod_{\alpha}  \exp\left(\int_0^t \mathcal{L}_{I\alpha} (t') dt'\right),
	\end{equation}
	which is possible because integrals of operators under time ordering commute.

    We define $\langle \cdot \rangle_{\alpha}$ as the expectation taken with respect to the initial state of bath $\alpha$ and set $\langle \mathcal{L}_{I\alpha}\rangle_{\alpha}=0$ for all $\alpha$ without loss of generality.
	Each bath can now be traced out independently using  $ \langle \exp (X) \rangle=\exp (\langle X^2\rangle/2  )$,
	true for any variable $X$ with a Gaussian distribution and zero mean. With this, and using idempotency of time ordering $\timo=\timo \timo$, we trace out the baths in Eq.~\eqref{eq:splitprop} to leave a dynamical map for the reduced system alone,
	\begin{align}\label{eq:eom}
		\Lambda_S(t) = & \timo  \prod_{\alpha}  \exp \left(\int_0^t\int_0^{t'} \langle \mathcal{L}_{I\alpha} (t') \mathcal{L}_{I\alpha} (t'') \rangle_{\alpha} dt' dt'' \right) \nonumber\\
	=&\timo \prod_{\alpha} \mathcal{F}_\alpha[s_\alpha^L,s_\alpha^R].
		\end{align}
	 Here we have enacted the time-ordering of the bath operators within the exponents in the first line, gaining a factor of 2, and have defined the system superoperator-valued influence functionals, $\mathcal{F}_\alpha[s_\alpha^L,s_\alpha^R]$.
	 
	 The system superoperator valued expectations of the interaction Liouvillians in Eq.~\eqref{eq:eom} are readily evaluated. We obtain
	 \begin{multline}\label{eq:LILI}
	    \langle \mathcal{L}_{I\alpha} (t') \mathcal{L}_{I\alpha} (t'') \rangle_{\alpha}=\\-s_\alpha^-(t')\left(s_\alpha^-(t'') C^\alpha_R(t'-t'')-is^+_\alpha(t'') C^\alpha_I(t'-t'')\right), 
	 \end{multline}
	where $s^\pm_\alpha=s^L_\alpha \pm s^R_\alpha$ and where $C^\alpha_R(t)$ and $C^\alpha_I(t)$ are the real and imaginary parts of
	\begin{multline}
	    C^\alpha(t)= \sum_q |g_{\alpha q}|^2\left( \cos(\omega_{\alpha q} t)\coth\left(\frac{\omega_{\alpha q}}{2T_\alpha}\right)-i \sin(\omega_{\alpha q} t)\right),
	\end{multline}
	the autocorrelation function of bath $\alpha$\footnote{A similar  derivation that we are describing here also holds for interactions of the form $H_I \sim s^\dagger a+sa^\dagger$, for both fermionic and bosonic environments. In both cases the only substantive difference is in the form of Eq.~\eqref{eq:LILI}.}.

	\subsection{Time-ordering using tensor contractions}
	Here we present the formalism required to cast Eq.~\eqref{eq:eom} into a tensor network, closely following that introduced in~\cite{strathearn2020modelling}. To  simplify our presentation we combine the influence functionals for each bath in Eq.~\eqref{eq:eom} into a single multi-bath influence function. Now we discretise the double integrals appearing in this influence functional onto a grid of  timesteps, $t_k = k \Delta$, where $0\geq k \geq N$. This allows us to write
	\begin{equation}\label{eq:discretize}
	 \Lambda_S(t_N) \approx \timo \prod_{k=0}^N \prod_{k'=0}^k \mathcal{I}_{k-k'}(t_k,t_{k'}),
	\end{equation}
	where 
	\begin{equation}
	    \mathcal{I}_{k-k'}(t_k,t_{k'})=  \exp \left(\sum_{\alpha} \langle \mathcal{L}_{I\alpha} (t_k) \mathcal{L}_{I\alpha} (t_{k'}) \rangle_{\alpha} \Delta^2 \right)
	\end{equation} 
	is a system superoperator-valued quantity that determines the influence of timestep $t_{k'}$ on $t_k$. Operators non-local in space generate spatial correlations when acting on a state and here we can equivalently think of the time non-local superoperators  $\mathcal{I}_{k-k'}(t_k,t_{k'})$  as generating non-Markovianity.
	
	The approximation made in Eq.~\eqref{eq:discretize} is equivalent to having made a Trotter splitting on the full system-baths propagator in Eq.~\eqref{eq:vnsol}, prior to tracing over the baths. The error incurred by the approximation then depends upon the order of the splitting. In practice we use a second-order splitting which incurs an error $\mathcal{O}(\Delta^3)$. We note that since the superoperators in the products in Eq.~\eqref{eq:discretize} do not commute in general, the products seem ambiguous. In fact, demanding consistency of Eq.~\eqref{eq:discretize} with making a Trotter splitting in Eq.~\eqref{eq:vnsol} fixes the order of multiplication of the superoperators to be from right to left with increasing $k$ and $k'$. Whether the $k$ product or the $k'$ product is performed first is irrelevant.

	We now need a prescription for enforcing time-ordering in Eq.~\eqref{eq:discretize}. If each $\mathcal{I}_{k-k'}(t_k,t_{k'})$ factorised into a product of time-local operators this would be trivial, since we could just arrange the overall product of operators in order, e.g.
    \begin{equation}\label{eq:TOex}
    \timo A(t_k)B(t_{k'}) = \left\{
    \begin{array}{lr}
    A(t_k)B(t_{k'}) & \text{if } t_k > t_{k'} \\
    B(t_{k'})A(t_k) & \text{if } t_{k'} > t_k
    \end{array}\right. .
    \end{equation}
    To illustrate how we enforce time ordering generally we now introduce some tensor network notation.  A rank-$n$ tensor is represented as a node with $n$ legs and contractions are indicated by connecting legs. Operators are rank-2 tensors so Eq.~\eqref{eq:TOex} has the following tensor network representation,
    \begin{equation}\label{eq:tenstimeord}
    \begin{tikzpicture}[baseline=(current  bounding  box.center),box/.style={rectangle,draw,very thick,minimum size=5mm,inner sep=0pt}]
    \node[box] (A) at (0,0) {$A$};
    \node (lb) [left=1mm of A] {$\timo \Bigg[$};
    \node[box] (B) [right=5mm of A] {$B$};
    \node (rb) [right=1mm of B] {$\Bigg] = $};

    \node (k) [above left=.1mm of B.north] {$j'$};
    \node (i) [above left=.1mm of A.north] {$j$};
    \node (l) [below left=.1mm of B.south] {$i'$};
    \node (j) [below left=.1mm of A.south] {$i$};
    
    \draw (A.north) -- ($(A)+(0,5mm)$);
    \draw ($(B)-(0,5mm)$) -- (B.south);
    \draw (A.south) -- ($(A)-(0,5mm)$);
    \draw (B.north) -- ($(B)+(0,5mm)$);
    
    \node[box] (C) at ($(rb.east)+ (8mm,7mm)$) {$A$};
    \node[box] (D) [right=5mm of C] {$B$};
    \node (t1>t2) [right=3mm of D] {if $t_k>t_{k'}$};
    
    \draw (C.north) -- ($(C)+(0,5mm)$);
    \draw ($(D)-(0,5mm)$) -- (D.south);
    \draw (C.south) -- ($(C)-(0,5mm)$) to [out=-90,in=90] ($(D)+(0,5mm)$) -- (D.north);
    
    \node[box] (E) at ($(rb.east)+ (8mm,-7mm)$) {$A$};
    \node[box] (F) [right=5mm of E] {$B$};
    \node (t2>t1) [right=3mm of F] {if $t_{k'}>t_k$};
    
    \draw (E.north) -- ($(E)+(0,5mm)$)to [out=90,in=-90] ($(F)-(0,5mm)$) -- (F.south);
    \draw ($(F)+(0,5mm)$) -- (F.north);
    \draw (E.south) -- ($(E)-(0,5mm)$);
    \draw [decorate,decoration={calligraphic brace,amplitude=10pt},xshift=-4pt,yshift=0pt,line width=1.25pt] ($(rb.east)+(4mm,-1.5cm)$) -- ($(rb.east)+(4mm,1.5cm)$);
    
    \end{tikzpicture}
    \end{equation}
    where $i$, $j$, $i'$, and $j'$ label the components of the matrices $A(t_k)$ and $B(t_{k'})$, i.e.  $[A(t_k)]_{i}^{j}$ and $[B(t_{k'})]_{i'}^{j'}$.
    On the left-hand side of Eq.~\eqref{eq:tenstimeord} we can write the two disconnected rank-2 tensors as a single rank-4 (4-leg) tensor by using a tensor product, $C(t_k,t_{k'})=A(t_k)\otimes B(t_{k'})$. The tensor $C(t_k,t_{k'})$ would then be represented by a single node,
    \begin{equation}\label{4leg}
    \begin{tikzpicture}[baseline=(current  bounding  box.center),box/.style={rectangle,draw,inner sep=2pt,align=center,fill=white, minimum height =2mm, minimum width=4mm}]
    \node (C) at (0,0) [draw, very thick,minimum width=1.6cm,minimum height=0.6cm] {$C$};
    \draw ($(C.north)+(5mm,0mm)$) -- ($(C.north)+(5mm,3mm)$);
    \draw ($(C.north)+(-5mm,0mm)$) -- ($(C.north)+(-5mm,3mm)$);
    \draw ($(C.south)+(-5mm,0mm)$) -- ($(C.south)+(-5mm,-3mm)$);
    \draw ($(C.south)+(5mm,0mm)$) -- ($(C.south)+(5mm,-3mm)$);
    \node (k) at ($(C.north)+(-6.5mm,3mm)$) {$j$};
    \node (i) at ($(C.north)+(3.5mm,3mm)$) {$j'$};
    \node (l) at ($(C.south)+(3.5mm,-3mm)$) {$i'$};
    \node (j) at ($(C.south)+(-6.5mm,-3mm)$) {$i$};

    \node (cin) [left=of C] {$[C(t_k,t_{k'})]_{ii'}^{jj'} \qquad \equiv$};
    \end{tikzpicture}
    \end{equation}
    where $[C(t_k,t_{k'})]_{ii'}^{jj'}=[A(t_k)]_{i}^j[B(t_{k'})]_{i'}^{j'}$.
    Comparing with Eq.~\eqref{eq:tenstimeord}, we see that enforcing time-ordering on this tensor would correspond to contracting two of the legs together; either the legs labelled by $i$ and $j'$ or those labelled by $j$ and $i'$. 
    
    % \begin{equation}\label{4leg}
    % \begin{tikzpicture}[baseline=(current  bounding  box.center),box/.style={rectangle,draw,very thick,minimum size=5mm,inner sep=0pt}]
    % \node[box] (C) at (0,0) {$C$};
    % \draw (C) -- ($(C)+(0,5mm)$);
    % \draw (C) -- ($(C)+(5mm,0)$);
    % \draw (C) -- ($(C)-(5mm,0)$);
    % \draw (C) -- ($(C)-(0,5mm)$);
    % \node (k) [above=1.5mm of C.north] {$i$};
    % \node (i) [right=1.5mm of C.east] {$k$};
    % \node (l) [left=1.5mm of C.west] {$l$};
    % \node (j) [below=1.5mm of C.south] {$j$};

    % \node (cin) [left=of C] {$[C(t_k,t_{k'})]_{ijkl} \equiv$};
    % \end{tikzpicture}.
    % \end{equation}
    
    We can formalise the use of a tensor product to combine time-local operators by way of a history Hilbert space~\cite{Cotler2018,Griffiths2002-GRICQT}.
    If $A(t_k)$ and $B(t_{k'})$ both act on a Hilbert space $\mathcal{H}$ spanned by a basis $\{\ket{e_i}\}$ then we can define a history space, $\mathcal{H}_{t_k}\otimes \mathcal{H}_{t_{k'}}$, spanned by $\{\ket{e_{i}}\otimes\ket{e_{i'}}\}$ on which their tensor product acts. Each subspace of the history space is identical to $\mathcal{H}$ and their corresponding bases are each identical to $\{\ket{e_i}\}$.
    In order to see how to write operators that depend on multiple time arguments in the history Hilbert space, let us consider an operator-valued function of the time-local operators $A(t_k)$ and $B(t_{k'})$, which we call $f(A(t_k)B(t_{k'}))$. This object exists in the familiar Hilbert space $\mathcal{H}$, and would have the usual matrix (i.e. rank-2 tensor) representation.
It can be converted into an operator that acts on the history space by simply changing the regular product into a tensor product as follows:
%
%
%
%For example if $f(t_k,t_{k'})=\exp(A(t_k)B(t_{k'}))$, which is a similar form to that of  $\mathcal{I}_{k-k'}(t_k,t_{k'})$, then we can define the history space operator
    \begin{align}
    D(t_k,t_{k'}) &= f(A(t_k)\otimes B(t_{k'})) \nonumber\\
    &= \sum_{i i' j j'} D_{ii'}^{jj'} \dyad{e_{i}}{e_{j}}\otimes \dyad{e_{i'}}{e_{j'}},
    \end{align} 
    where 
    \begin{equation}
    D_{i i'}^{jj'} = \left(\bra{e_i}\otimes\bra{e_{i'}}\right) D(t_k,t_{k'})\left(\ket{e_{j}}\otimes\ket{e_{j'}}\right).
    \end{equation}
    The operator $D(t_k,t_{k'})$ is a rank-4 tensor which is represented in the same way as $C(t_k,t_{k'})$ in Eq.~\eqref{4leg}. Crucially, to enforce time ordering on $D(t_k,t_{k'})$, i.e. to evaluate $\timo f(A(t_k)B(t_{k'})) $, we simply connect together two of its legs, exactly as we did for $C(t_k,t_{k'})$.
    
    The expression we wish to time order in Eq.~\eqref{eq:discretize} is a function of superoperators acting at $N+1$ different points in time. Thus, we construct the history space, $\mathfrak{L}_H$, as a product of $N+1$ copies of the Liouville space, $\mathfrak{L}$, in which the vectorised system density matrix, $\sket{\rho_S},$ is defined. If the full discretised influence functional is converted to a superoperator on $\mathfrak{L}_H$ it is represented diagrammatically as a rank-$2(N+1)$ tensor, the influence tensor, where each of the time points gives rise to a pair of legs. Time ordering is then enforced by connecting pairs of legs arising from consecutive time points, leaving two unconnected which are the two legs of the resulting dynamical map which is a rank-2 tensor. So if we let $N=3$ then Eq.~\eqref{eq:discretize} has the following tensor network representation
    \begin{equation}\label{Nleg}
    \begin{tikzpicture}[baseline=(current  bounding  box.center),box/.style={rectangle,draw,inner sep=2pt,align=center,fill=white, minimum height =2mm, minimum width=4mm}]
    \node (C) at (0,0) [draw, very thick,minimum width=3cm,minimum height=0.6cm] {};
    \draw ($(C.north)+(12mm,0mm)$) -- ($(C.north)+(12mm,2.5mm)$);
    \draw ($(C.south)+(12mm,0mm)$) -- ($(C.south)+(12mm,-2.5mm)$);
    \draw ($(C.north)+(4mm,0mm)$) -- ($(C.north)+(4mm,2.5mm)$);
    \draw ($(C.south)+(4mm,0mm)$) -- ($(C.south)+(4mm,-2.5mm)$);
    \draw ($(C.north)+(-4mm,0mm)$) -- ($(C.north)+(-4mm,2.5mm)$);
    \draw ($(C.south)+(-4mm,0mm)$) -- ($(C.south)+(-4mm,-2.5mm)$);
    \draw ($(C.north)+(-12mm,0mm)$) -- ($(C.north)+(-12mm,2.5mm)$);
    \draw ($(C.south)+(-12mm,0mm)$) -- ($(C.south)+(-12mm,-2.5mm)$);
    
    \draw ($(C.north)+(12mm,0mm)$) -- ($(C.north)+(12mm,2.5mm)$) to [out=90,in=-90] ($(C.south)+(4mm,-2.5mm)$)-- ($(C.south)+(4mm,0mm)$) ;
    \draw ($(C.north)+(4mm,0mm)$) -- ($(C.north)+(4mm,2.5mm)$) to [out=90,in=-90] ($(C.south)+(-4mm,-2.5mm)$)-- ($(C.south)+(-4mm,0mm)$) ;
    \draw ($(C.north)+(-4mm,0mm)$) -- ($(C.north)+(-4mm,2.5mm)$) to [out=90,in=-90] ($(C.south)+(-12mm,-2.5mm)$)-- ($(C.south)+(-12mm,0mm)$) ;
    
    %\draw ($(C.north)+(4mm,2.5mm)$) -- ($(C.south)+(12mm,-2.5mm)$);
    %\draw ($(C.north)+(-4mm,2.5mm)$) -- ($(C.south)+(4mm,-2.5mm)$);
    %\draw ($(C.north)+(-12mm,2.5mm)$) -- ($(C.south)+(-4mm,-2.5mm)$);
    
    \node (cin) [left=of C] {$=$};
    \node (lamb) [draw,very thick, left=of cin] {$\Lambda_S$};
    
    \draw ($(lamb.north)$) -- ($(lamb.north)+(0mm,2.5mm)$);
    
    \draw ($(lamb.south)$) -- ($(lamb.south)+(0mm,-2.5mm)$);
    \end{tikzpicture}
    \end{equation}
    where each pair of legs on opposing sides of the influence tensor are those arising from a single subspace of $\mathfrak{L}_H$ and we have arranged the legs so that right most pair correspond to $\mathfrak{L}_{t_0}$, the next pair along correspond to $\mathfrak{L}_{t_1}$, and so on.
    
    Rather than representing Eq.~\eqref{eq:discretize} as a contraction of a single, large tensor in $\mathfrak{L}_H$, we note that each individual $\mathcal{I}_{k-k'}(t_k,t_{k'})$  can be represented either as a rank-4 tensor in the subspace $\mathfrak{L}_{t_k} \otimes \mathfrak{L}_{t_{k'}}$ (for $k\ne k'$) or as a rank-2 tensor in the subspace $\mathfrak{L}_{t_k}$ (for $k = k'$).   This means that the rank-$2(N+1)$ influence tensor can be decomposed into a network of rank-2 and rank-4 tensors.     For example, the string of tensor contractions corresponding to $N=3$ and a fixed $k'=0$ in Eq.~\eqref{eq:discretize} would be represented as
	\begin{equation}\label{network:row}
	\begin{tikzpicture}[baseline=(current  bounding  box.center),box/.style={rectangle,draw,very thick,minimum size=7mm,inner sep=0pt}]
	
	\matrix[row sep=4mm,column sep=4mm] (m) at (0,0) {
		\node[box] (I30) {$\mathcal{I}_3$}; & \node[box] (I31) {$\mathcal{I}_2$}; & \node[box] (I32) {$\mathcal{I}_1$}; & \node[box] (I33) {$\mathcal{I}_0$};\\};
	
	\graph{(I30) -- (I31) -- (I32) -- (I33);};
	
	\foreach \x in {I30,I31,I32}
	{
		\draw (\x.north) -- ($(\x.north)+(0,2mm)$);
		\draw (\x.south) -- ($(\x.south)-(0,2mm)$);
	}
	
	\draw (I33.south) -- ($(I33.south)-(0,2mm)$);
	
	\draw (I30.west) -- ($(I30.west)-(2mm,0)$);

	\end{tikzpicture},
	\end{equation}
	where contractions occur between the legs which act commonly on $\mathfrak{L}_{t_0}$. The full influence tensor in Eq.~\eqref{eq:discretize} with $N=3$, prior to time ordering, would then be written as
	\begin{equation}\label{network:full}
	\begin{tikzpicture}[baseline=(current  bounding  box.center),box/.style={rectangle,draw,very thick,minimum size=7mm,inner sep=0pt}]
	
	\matrix[row sep=4mm,column sep=4mm] (m) at (0,0) {
		\node[box] (I33) {$\mathcal{I}_0$}; &; &; &;\\
		 \node[box] (I32) {$\mathcal{I}_1$}; & \node[box] (I22) {$\mathcal{I}_0$}; &; &;  \\
		\node[box] (I31) {$\mathcal{I}_2$}; & \node[box] (I21) {$\mathcal{I}_1$}; & \node[box] (I11) {$\mathcal{I}_0$}; &;  \\
		\node[box] (I30) {$\mathcal{I}_3$}; & \node[box] (I20) {$\mathcal{I}_2$}; & \node[box] (I10) {$\mathcal{I}_1$}; & \node[box] (I00) {$\mathcal{I}_0$}; \\};
	
	\graph{[edges={white!50!green,line width=5pt}] (I20) -- (I21) -- (I22) -- (I32);};
	\draw[line width=5pt,white!50!green] (I32.west) -- ($(I32.west)-(2mm,0)$);
	\draw[line width=5pt,white!50!green] (I20.south) -- ($(I20.south)-(0,2mm)$);

	\graph{ (I00) -- (I10) -- 
		(I20) -- (I30);
		(I10) -- (I11) -- (I21) -- (I31);
		(I20) -- (I21) -- (I22) -- (I32);
		(I30) -- (I31) -- (I32) -- (I33);};

	\foreach \x in {I00,I10,I20,I30}
	\draw (\x.south) -- ($(\x.south)-(0,2mm)$);
	
	\foreach \x in {I30,I31,I32,I33}
	\draw (\x.west) -- ($(\x.west)-(2mm,0)$);

	\end{tikzpicture}
	\end{equation}
	where we have highlighted in green a single string of contractions occurring through the $\mathfrak{L}_{t_2}$ subspace of $\mathfrak{L}_H$. The dangling legs along at the bottom of (\ref{network:full}) are the same as those on the bottom of the influence tensor in Eq.~\eqref{Nleg}, while the legs along the left of (\ref{network:full}) going from bottom to top are the same as those on the top in Eq.~\eqref{Nleg} going from right to left. The dimension of the bonds and the dangling legs in this network is $d^2$.
	
	In the case where $[s_\alpha,s_{\alpha'}]=0$ for all $\alpha,\alpha'$ the nodes in (\ref{network:full}) can be expanded in terms of the complete, shared eigenbasis of the system operators $s_\alpha$. Upon enforcing time ordering it is then possible to recover the standard path sum representation of the influence functional \cite{strathearn2020modelling} and the network can be contracted efficiently using the original TEMPO algorithm \cite{strathearn2017efficient}, or by other means \cite{jorgensen2019exploiting}. In the general case that the $s_\alpha$ do not commute enforcing time ordering upon and contracting (\ref{network:full}) efficiently has so far not been possible. We present a method that achieves this in the following section.
	
	It should be pointed out that the influence tensor we have derived above is in fact a process tensor \cite{jorgensen2019exploiting,pollock2018nonmarkovian,milz2020genuine}. A process tensor is a mapping from the space of possible inputs to a system at a discrete set of times, for example a measurement or a preparation, to output states. Therefore not only can (\ref{network:full}) be used to calculate the dynamical map for the system but also all of its correlation functions \cite{Fux2021,jorgensen2019exploiting}. Recently it has been shown that this further allows the calculation of all correlation functions of the baths as well \cite{Gribben2021}. Thus, the tensor representation for the influence functional introduced here provides a way to completely characterise the full system-baths dynamics.

	\subsection{Matrix product form of the influence tensor}
	Here we show how the influence tensor derived in the previous section can be efficiently put into the form of a matrix product operator (MPO). The first step is to convert each row of tensors in (\ref{network:full}) individually into an MPO. 	This can be done via singular value decompositions (SVD) using the following routine:
	\begin{equation}\label{network:movinglegs}
	\begin{tikzpicture}[baseline=(current  bounding  box.center),box/.style={rectangle,draw,very thick,minimum size=7mm,inner sep=0pt}]
	\node[box] (A) at (0,0) {};
	
    \node[left= 1cm of A] (step1) {(a)};

	\node[box] (B) [right=4mm of A] {};
	\draw[line width=4pt,white!80!blue] ($(A.west)-(2mm,0)$) -- (A);
	\draw ($(A.west)-(2mm,0)$) -- (A);
	\draw (A) -- (B) node[midway] (AB) {} -- ($(B.east)+(2mm,0)$);
	\foreach \x in {A,B}
	\draw ($(\x.south)-(0,2mm)$)--(\x)--($(\x.north)+(0,2mm)$);
	
	\draw [->,line width = .5mm] ($(AB)-(0,8mm)$) -- ($(AB)-(0,12mm)$);
	
	\node[box] (A1) [below=12mm of A] {};
	
	\node[left= 1cm of A1] (step1) {(b)};
	
	\node[box] (B1) [right=4mm of A1] {};
	\draw[line width=4pt,white!80!blue] ($(A1.east)+(0mm,1mm)$)--($(A1.east)+(2mm,1mm)$);
	\draw ($(A1.east)+(0mm,1mm)$)--($(A1.east)+(2mm,1mm)$);
	\draw (A1) -- (B1) node[midway] (AB1) {} -- ($(B1.east)+(2mm,0)$);
	\foreach \x in {A1,B1}
	\draw ($(\x.south)-(0,2mm)$)--(\x)--($(\x.north)+(0,2mm)$);
	\draw[dashed] (A1.north east) to [out=-150,in=150] (A1.south east);
	
	\draw [->,line width = .5mm] ($(AB1)-(0,8mm)$) -- ($(AB1)-(0,12mm)$) node[right=6mm,midway] {SVD};
	
	\node[box] (A2) [below=12mm of A1] {$U$};
	
	\node[left= 1cm of A2] (step1) {(c)};
	
	\node[diamond,draw,minimum size=6mm,inner sep=0mm] (B2) [right=2mm of A2] {$s$};
	\node[box] (C2) [right=2mm of B2] {$V^\dagger$};
	\node[box] (D2) [right=4mm of C2] {};
	\draw[line width=4pt,white!80!blue] ($(C2.east)+(0mm,1mm)$)--($(C2.east)+(2mm,1mm)$);
	\draw ($(C2.east)+(0mm,1mm)$)--($(C2.east)+(2mm,1mm)$);
	\foreach \x in {A2,D2}
	\draw ($(\x.south)-(0,2mm)$)--(\x)--($(\x.north)+(0,2mm)$);
	\draw (A2) -- (B2) node[midway] (AB2) {} -- (C2) -- (D2) -- ($(D2.east)+(2mm,0)$);
	\node[rectangle,fit=(B2)(D2),draw,dashed,minimum height = 12.5mm] {};
	
    \draw [->,line width = .5mm] ($(AB2)-(0,8mm)$) -- ($(AB2)-(0,12mm)$) node[right=6mm,midway] {contract};
	
	\node[box] (A3) [below=12mm of A2] {};
	
	\node[left= 1cm of A3] (step1) {(d)};
	
	\node[box] (B3) [right=4mm of A3] {};
	\draw[line width=4pt,white!80!blue] ($(B3.east)+(0mm,1mm)$)--($(B3.east)+(2mm,1mm)$);
	\draw ($(B3.east)+(0mm,1mm)$)--($(B3.east)+(2mm,1mm)$);
	\draw (A3) -- (B3) -- ($(B3.east)+(2mm,0)$);
	\foreach \x in {A3,B3}
	\draw ($(\x.south)-(0,2mm)$)--(\x)--($(\x.north)+(0,2mm)$);

	\end{tikzpicture}.
	\end{equation} 
	Here, we start with a pair of rank-4 tensors connected by a single leg as shown in (a). The desired outcome is shown in (d) with the dimension corresponding to the blue highlighted leg being stored by the right-hand tensor. This corresponds to an alternate decomposition of the rank-6 tensor computed from contracting the bond in (a). The steps in \eqref{network:movinglegs} show how to switch to this representation without constructing the full tensor. In (b) we indicate with a dashed line that the legs of the tensor are partitioned into two groups; these are combined to form a matrix. We then perform an SVD on this matrix to arrive at step (c). Moving from (c) to (d) we contract the sub-network indicated by the dashed box and in doing so arrive at the desired representation.
	
	%This procedure can alternatively be done analytically without the use of SVDs and without increasing the bond dimension, $d^2$, between the two tensors in the first and final steps of (\ref{network:movinglegs}). However by using an SVD it is possible to reduce the bond dimension which makes further contractions computationally easier. 
	
	We can now use (\ref{network:movinglegs}) to move the legs on the left of each row in (\ref{network:full}), which puts each row in MPO form. For the bottom row this looks like:
	\begin{equation}\label{network:rowmovingleg}
	\begin{tikzpicture}[baseline=(current  bounding  box.center),box/.style={rectangle,draw,very thick,minimum size=7mm,inner sep=0pt}]
	
	\matrix[row sep=4mm,column sep=4mm] (m) at (0,0) {
		\node[box] (I30) {$\mathcal{I}_3$}; & \node[box] (I20) {$\mathcal{I}_2$}; & \node[box] (I10) {$\mathcal{I}_1$}; & \node[box] (I00) {$\mathcal{I}_0$};\\};
	
	\graph{(I30) -- (I20) -- (I10) -- (I00);};
	
	\foreach \x in {I30,I20,I10}
	{
		\draw (\x.north) -- ($(\x.north)+(0,2mm)$);
		\draw (\x.south) -- ($(\x.south)-(0,2mm)$);
	}
	
	\draw (I00.south) -- ($(I00.south)-(0,2mm)$);
	
	\draw (I30.west) -- ($(I30.west)-(2mm,0)$);
	
	%\draw[->] plot [smooth,tension=0.01] coordinates {($(I30.west)-(3mm,0)$) ($(I30)-(5mm,5mm)$) ($(I30.south)-(0,3mm)$) ($(I31.south)-(0,3mm)$) ($(I32.south)-(0,3mm)$) ($(I33.south)-(0,2mm)$)};
	
	\draw[->,dashed] ($(I30.west)-(4mm,0)$) to[out=170,in=150] ($(I00.north)+(0,2mm)$);
	
	\draw [->,line width = .5mm] ($(m)-(0,8mm)$) -- ($(m)-(0,12mm)$);
	
	\matrix[row sep=4mm,column sep=4mm] (m2) [below=12mm of m] {
		\node[box] (I30) {$\mathcal{I}_3$}; & \node[box] (I20) {$\mathcal{I}_2$}; & \node[box] (I10) {$\mathcal{I}_1$}; & \node[box] (I00) {$\mathcal{I}_0$};\\};
	
	\graph{(I30) -- (I20) -- (I10) -- (I00);};
	
	\foreach \x in {I30,I20,I10,I00}
	{
		\draw (\x.north) -- ($(\x.north)+(0,2mm)$);
		\draw (\x.south) -- ($(\x.south)-(0,2mm)$);
	}

	\end{tikzpicture}.
	\end{equation}
    Having done this for each row, the full influence tensor can now be computed by iteratively contracting together the product of resulting MPOs, using SVDs to maintain an efficient representation:
    	\begin{equation}\label{network:fulllegsmoved}
	\begin{tikzpicture}[baseline=(current  bounding  box.center),box/.style={rectangle,draw,very thick,minimum size=7mm,inner sep=0pt}]
	
	\matrix[row sep=4mm,column sep=4mm] (m) at (0,0) {
		\node[box] (I33) {$\mathcal{I}_0$}; &; &; &;\\
		 \node[box] (I32) {$\mathcal{I}_1$}; & \node[box] (I22) {$\mathcal{I}_0$}; &; &;  \\
		\node[box] (I31) {$\mathcal{I}_2$}; & \node[box] (I21) {$\mathcal{I}_1$}; & \node[box] (I11) {$\mathcal{I}_0$}; &;  \\
		\node[box] (I30) {$\mathcal{I}_3$}; & \node[box] (I20) {$\mathcal{I}_2$}; & \node[box] (I10) {$\mathcal{I}_1$}; & \node[box] (I00) {$\mathcal{I}_0$}; \\};

	\graph{ (I00) -- (I10) -- 
		(I20) -- (I30);
		(I10) -- (I11) -- (I21) -- (I31);
		(I20) -- (I21) -- (I22) -- (I32);
		(I30) -- (I31) -- (I32) -- (I33);};

	\foreach \x in {I30,I20,I10,I00}
	\draw (\x.south) -- ($(\x.south)-(0,2mm)$);
	
	\foreach \x in {I00,I11,I22,I33}
	\draw (\x.north) -- ($(\x.north)+(0,2mm)$);

	\matrix[row sep=4mm,column sep=4mm] (m2) [below=12mm of m] {
		\node[box] (I30) {}; & \node[box] (I31) {}; & \node[box] (I32) {}; & \node[box] (I33) {};\\};
	
	\graph{(I30) -- (I31) -- (I32) -- (I33);};
	
	\foreach \x in {I30,I31,I32,I33}
	{
		\draw (\x.north) -- ($(\x.north)+(0,2mm)$);
		\draw (\x.south) -- ($(\x.south)-(0,2mm)$);
	}

	\draw [->,line width = .5mm] ($(m2)+(0,12mm)$) -- ($(m2)+(0,8mm)$);
	
	\end{tikzpicture}.
	\end{equation}
	
	The result is a MPO representation of the influence tensor, where the vertical legs on each node correspond to the input (below) and output (above) legs of a single subspace of the full history space. That is, the rightmost node takes a state at $t_0$ as an input from below and outputs a state at $t_1$. The next node to the left would take this state from $t_1$ to $t_2$ and so on. Enforcing time ordering on the influence tensor to obtain the dynamical map, as in Eq.~\eqref{Nleg}, is now straightforward. Doing this and acting the dynamical map on an initial state to obtain a final state at time $t_4$ would look like
	\begin{equation}\label{eq:eomtimeordered}
	\begin{tikzpicture}
	[baseline=(current  bounding  box.center),box/.style={rectangle,draw,very thick,minimum size=7mm,inner sep=0pt}]
	\matrix[row sep=4mm,column sep=4mm] (m) at (0,0) {
		\node[box] (I30) {}; &
		\node[box] (I20) {}; &
		\node[box] (I10) {}; &
		\node[box] (I00) {};\\};
	
	\node[circle,draw,fill=red] (rho0) [below=4mm of I00] {};
	\draw (rho0) -- (I00);

	\draw (I00) -- (I10) -- (I20) -- (I30);
	
	\draw (I00) -- ($(I00.north)+(0,2mm)$) to [out=90,in=-90] ($(I10.south)-(0,2mm)$) -- (I10) 
	-- ($(I10.north)+(0,2mm)$) to [out=90,in=-90] ($(I20.south)-(0,2mm)$) -- (I20)
	-- ($(I20.north)+(0,2mm)$) to [out=90,in=-90] ($(I30.south)-(0,2mm)$) -- (I30) -- ($(I30.north)+(0,2mm)$);
	
	%\node (rhotext) at [right=2mm of I30] {$\sket{\rho(t_4)}$};
	\node (rhotext) [above=2mm of I30] {$\sket{\rho(t_4)}$};
	
	\end{tikzpicture}
	\end{equation}
	where the (rank-1) initial state is represented as a red circle.

	There is one further step we can take to make the calculation more efficient. Instead of combining the individual influence functionals for each bath and then discretising the results, as in Eq.~\eqref{eq:discretize}, we can discretise each $\mathcal{F}_\alpha[s_\alpha^L,s_\alpha^R]$ in Eq.~\eqref{eq:eom} separately, and obtain a MPO influence tensor for each of them individually using the procedure outlined above. The single MPO used to find the dynamical map in (\ref{eq:eomtimeordered}) is then calculated by contracting together the MPOs for each $\alpha$. In the case that there are just two baths, $\alpha_1$ and $\alpha_2$, the influence tensor MPO used in (\ref{eq:eomtimeordered}) would be written:
	\begin{equation}\label{eq:eomnetwork}
	\begin{tikzpicture}
	[baseline=(current  bounding  box.center),box/.style={rectangle,draw,very thick,minimum size=7mm,inner sep=0pt}]
	\matrix[row sep=4mm,column sep=4mm] (m) at (0,0) {
		\node[box] (I30) {}; &
		\node[box] (I20) {}; &
		\node[box] (I10) {}; &
		\node[box] (I00) {};\\};

	\matrix[row sep=4mm,column sep=4mm] (m2) [above=5mm of m] {
		\node[box] (I301) {}; &
		\node[box] (I201) {}; &
		\node[box] (I101) {}; &
		\node[box] (I001) {};\\};
	
	\node[right=4mm of I00] () {$\mathcal{F}_{\alpha_1}$};
	\node[right=4mm of I001] () {$\mathcal{F}_{\alpha_2}$};
	
	\graph{ (I00) -- (I10) -- 
		(I20) -- (I30);
		(I001) -- (I101) -- 
		(I201) -- (I301);
	};
	
	\foreach \x in {I00,I10,I20,I30}
	{
		\node[circle,draw] (freeprop\x) [above=4mm of \x1] {};
		\draw ($(\x.south)-(0,2mm)$)-- (\x) -- (\x1) -- (freeprop\x) -- ($(freeprop\x.north)+(0,2mm)$);
	}
	
	\node[right=4mm of freepropI00] () {$\mathrm{e}^{\mathcal{L}_S \Delta t}$};
	
	\end{tikzpicture}.
	\end{equation}
	Here we have returned from the system interaction picture to the Schr\"odinger picture. The result of this is that the nodes in the bulk of the network are now defined in terms of time-independent system superoperators, e.g. $s^L_\alpha(t)=s^L_\alpha$, and free system propagators, rank-2 tensors represented as white circles, are applied in each subspace of the history space. The appearance of the free system propagators can be understood using the language of process tensors: in between the periods of time where the system interacts with the baths (the process) the operations we perform on the system are simply those of free evolution. 
	
	Calculating the influence tensor MPOs for each bath individually and then combining them does not result in a final MPO with lower bond dimension than if we had first combined the bath influence functionals and then directly calculated the full multi-bath MPO. This is to be expected since each method of calculation results in the same final MPO, up to the errors associated with Trotterization and SVD truncation. However, we find that the individual MPO calculations for each bath are computationally less intensive than the alternative so we can do multiple easier calculations rather than a single intensive one.
	
	The method we have introduced here can be taken even further by breaking down the MPOs for each bath into a series of MPOs representing the individual modes that make up the bath. This approach is similar in spirit to other recent works~\cite{cygorek2021numericallyexact,ye2021constructing} where the process tensor can be constructed in a purely numerical manner, even when the environment is not Gaussian and the influence functional cannot be calculated analytically in closed form. Finally, we stress that process tensors calculated using different approaches for each environment can be combined like this, provided time is discretized in the same way. If the system coupling operators for different environments are related by a unitary transformation then the resulting process tensors can be transformed in the same way.

	\subsection{Computing all dynamical maps efficiently}
	The methodology we have presented so far enables us to calculate the dynamical map between two times, $t_0$ and $t_N$. In this section we explain how to extract dynamical maps between $t_0$ and all times $t<t_N$ during a single computation and give details on how we maximise the efficiency of the method. Generally we aim to minimise the size of the tensors that must be stored while computing the dynamical maps. There are two steps we take towards this: we enforce time ordering on-the-fly during construction of the full multi-bath influence tensor and we keep contributions from each single-bath influence functional separate until absolutely necessary. We found this second point particularly helpful since combining influence tensors for non-commuting environments leads to a significant increase in bond dimension. The influence tensor is really a tensor representation of the integrand of a path sum which contains all possible trajectories the systems takes while interacting with its environment. Contracting together two single-bath influence tensors combines two sets of such trajectories. If the couplings to these baths commute then many of these trajectories will be degenerate and the bond dimension remains low, otherwise a larger bond dimension will be needed to store the combination of trajectories.
	
	To recover a dynamical map from time $t_0$ to some time $t_n$ from an influence tensor constructed with a final time $t_N>t_n$ we need to discard from the tensor all correlations between pairs of time points that straddle the time $t_n$. We do this using a (normalised) vectorized identity matrix, $d^{-1/2}\sket{\mathbf{1}}$, a rank-1 tensor which we call a trace cap since $\langle \langle \mathbf{1}|O\rangle \rangle=\mathrm{Tr}[O]$ for any vectorized system operator $O$. The trace cap is a null vector of any commutator superoperator, e.g $s_\alpha^-\sket{\mathbf{1}}=0$, and upon inspection of the form of Eq.~\eqref{eq:LILI}, which is the exponent of an influence funcional, we can see that the result of applying the trace caps to the influence tensor at time $t_k$ is to erase any temporal correlation that is generated between this time and all earlier times $t<t_k$. 
	In practice we use the trace cap as follows:
	\begin{equation}\label{eq:dynmapcalc}
	\begin{tikzpicture}
	[baseline=(current  bounding  box.center),box/.style={rectangle,draw,very thick,minimum size=7mm,inner sep=0pt}]
	\matrix[row sep=4mm,column sep=4mm] (m) at (0,0) {
		\node[box] (I30) {}; &
		\node[box] (I20) {}; &
		\node[box] (I10) {}; &
		\node[box] (I00) {}; \\};

	\draw (I00) -- (I10) -- (I20) -- (I30);
	
	\draw (I00) -- ($(I00.north)+(0,2mm)$) to [out=90,in=-90] ($(I10.south)-(0,2mm)$) -- (I10) 
	-- ($(I10.north)+(0,2mm)$) to [out=90,in=-90] ($(I20.south)-(0,2mm)$) -- (I20)
	-- ($(I20.north)+(0,2mm)$);
	\draw ($(I30.south)-(0,2mm)$) node (I30l) {} -- (I30) -- ($(I30.north)+(0,2mm)$) node (I30r) {};
	\draw ($(I00.south)-(0,2mm)$)--(I00);
	\filldraw[color=black,fill=green!50!black,very thick] ($(I30l)+(3mm,0mm)$) arc [radius={3mm},start angle={0},end angle={-180}] node[] (capl) {};
	\draw[very thick] ($(I30l)-(3mm,0mm)$)--($(I30l)+(3mm,0mm)$);
	
	\filldraw[color=black,fill=green!50!black,very thick] ($(I30r)+(3mm,0mm)$) arc [radius={3mm},start angle={0},end angle={180}] node[] (capr) {};
	\draw[very thick] ($(I30r)-(3mm,0mm)$)--($(I30r)+(3mm,0mm)$);

	\node (captext) [above=6mm of I30.north] {Trace cap};
	
	\end{tikzpicture},
	\end{equation}
	where the trace caps are shown as green semicircles and we have extracted the dynamical map up to time $t_3$ from an influence tensor with $N=3$. Trace caps can be used in this way either on the full multi-bath influence tensor, as we did above, or on the single-bath influence tensors. For instance, if we write the influence tensor in \eqref{eq:dynmapcalc} explicitly as a combination of two single-bath influence tensors and free system propagators the trace caps are applied as follows:
	\begin{equation}\label{eq:twoenvnetwork}
	\begin{tikzpicture}
	[baseline=(current  bounding  box.center),box/.style={rectangle,draw,very thick,minimum size=7mm,inner sep=0pt}]
	
	\matrix[row sep=4mm,column sep=4mm] (m4) at (0,0) {
		\node[box] (I30b1) {}; &
		\node[box] (I20b1) {}; &
		\node[box] (I10b1) {}; &
		\node[box] (I00b1) {};\\};
	\matrix[row sep=4mm,column sep=4mm] (m3) [below=1cm of m4] {
		\node[box] (I30b) {}; &
		\node[box] (I20b) {}; &
		\node[box] (I10b) {}; &
		\node[box] (I00b) {};\\};

	\graph{ (I00b) -- (I10b) -- 
		(I20b) -- (I30b);
		(I00b1) -- (I10b1) -- 
		(I20b1) -- (I30b1);
	};
	
	\foreach \x in {I00b,I10b,I20b}
	{
		\node[circle,draw] (freeprop\x) [above=4mm of \x1] {};
		\draw ($(\x.south)-(0,2mm)$) node (\x l) {}-- (\x) -- (\x1) -- (freeprop\x) -- ($(freeprop\x.north)+(0,2mm)$) node (\x r) {};
	}
	
	\draw ($(I30b.south)-(0,2mm)$) node (I30bl) {} -- (I30b);
	\draw ($(I30b.north)+(0,2mm)$) node (I30br) {} -- (I30b);
	\draw ($(I30b1.south)-(0,2mm)$) node (I30b1l) {} -- (I30b1);
	\draw ($(I30b1.north)+(0,2mm)$) node (I30b1r) {} -- (I30b1);
	
	\draw ($(freepropI00b.north)+(0,2mm)$) to [out=90,in=-90] ($(I10b.south)-(0,2mm)$);
	\draw ($(freepropI10b.north)+(0,2mm)$) to [out=90,in=-90] ($(I20b.south)-(0,2mm)$);

	\filldraw[color=black,fill=green!50!black,very thick] 
	($(I30bl)+(3mm,0mm)$) arc [radius={3mm},start angle={0},end angle={-180}] node[] (capl) {};
	\draw[very thick] ($(I30bl)-(3mm,0mm)$)--($(I30bl)+(3mm,0mm)$);
	
	\filldraw[color=black,fill=green!50!black,very thick] ($(I30br)+(3mm,0mm)$) arc [radius={3mm},start angle={0},end angle={180}] node[] (capl) {};
	\draw[very thick] ($(I30br)-(3mm,0mm)$)--($(I30br)+(3mm,0mm)$);
	
	\filldraw[color=black,fill=green!50!black,very thick] ($(I30b1l)+(3mm,0mm)$) arc [radius={3mm},start angle={0},end angle={-180}] node[] (capl) {};
	\draw[very thick] ($(I30b1l)-(3mm,0mm)$)--($(I30b1l)+(3mm,0mm)$);
	
	\filldraw[color=black,fill=green!50!black,very thick] ($(I30b1r)+(3mm,0mm)$) arc [radius={3mm},start angle={0},end angle={180}] node[] (capl) {};
	\draw[very thick] ($(I30b1r)-(3mm,0mm)$)--($(I30b1r)+(3mm,0mm)$);

	\end{tikzpicture}.
	\end{equation}
	
	To see how we enforce time ordering on-the-fly, note that all the information required to calculate a dynamical map to time $t_n<t_N$ is contained in the first $n$ rows from the bottom in the influence tensor in the first diagram of (\ref{network:fulllegsmoved}). So, if we want to calculate the dynamical map from $t_0$ to $t_2$ we first need to contract together the first two rows, enforce time ordering only between the $t_0$ and $t_1$ nodes,
	\begin{equation}\label{eq:dynmapcalc2}
	\begin{tikzpicture}
	[baseline=(current  bounding  box.center),box/.style={rectangle,draw,very thick,minimum size=7mm,inner sep=0pt}]
	\matrix[row sep=4mm,column sep=4mm] (m) at (0,0) {
		
		\node[box] (I10) {}; &
		\node[box] (I00) {};\\};

	\draw ($(I10.west)$) -- ($(I10.west)-(2mm,0)$);

	\draw (I00) -- (I10) ;
	
	\draw ($(I00.south)-(0,2mm)$) -- (I00) -- ($(I00.north)+(0,2mm)$) to [out=90,in=-90] ($(I10.south)-(0,2mm)$) -- (I10) 
	-- ($(I10.north)+(0,2mm)$);
    
    \matrix[row sep=4mm,column sep=4mm] (m2) [below=8mm of m] {
		\node[box] (I30) {};\\};

	\draw ($(I30.south)-(0,2mm)$) -- (I30) -- ($(I30.north)+(0,2mm)$);
	\draw ($(I30.west)$) -- ($(I30.west)-(2mm,0)$);
	
	\draw[->,line width = .5mm] ($(m)-(0,7mm)$) -- ($(m)-(0,11mm)$);
	
	\end{tikzpicture}
	\end{equation}
	then apply trace caps to the rest of the MPO. Before applying the trace caps we store a copy of the MPO so that it can be used for further computation but the time ordering step here is permanent, since the legs that were contracted in this step, \eqref{eq:dynmapcalc2}, have no further use in the calculation. Thus, the MPO we use to continue the calculation has only $N$ nodes, one fewer than the initial MPO we began with. Continuing to obtain the dynamical map to $t_3$, we contract our stored MPO with the third row in (\ref{network:fulllegsmoved}), use Eq.~\eqref{eq:dynmapcalc2}, then apply trace caps. After this we have stored an MPO that has $N-1$ nodes. In this way the MPO we store gets smaller as we proceed with the calculation.
	
	Similarly to how we enforce time ordering as described above, we also combine single-bath influence tensors on-the-fly. To calculate the dynamical map from $t_0$ to $t_n$, we need to contract $n$ rows of (\ref{network:fulllegsmoved}) for each single-bath influence tensor, then we only need to combine the $n$ nodes of the resulting MPOs corresponding to times $t\leq t_{n-1}$ before applying time ordering on these nodes. After storing the the remaining nodes of these separate MPOs, those for $t>t_{n-1}$, trace caps are used as in (\ref{eq:twoenvnetwork}) to obtain the dynamical map. In practice we did not contract the initial state with the influence tensor while calculating the dynamical maps. This allowed us to optimise our choice of initial state in order to find the steady state in as few time steps as possible, which, since we considered two baths in our calculations, meant we also had to store a rank-4 tensor during computations. In summary, the following schematic tensor network shows which tensors we have stored after calculating the first $k$ dynamical maps:
	\begin{equation}
	\begin{tikzpicture}[baseline=(current  bounding  box.center),box/.style={rectangle,draw,inner sep=2pt,align=center,fill=white}]
	\node[box] (base) at (0,0) {Tensor carrying dynamical\\information up to $t_k$};
	\draw ($(base.west)-(5mm,0)$) node (init) {} --(base.west);
	
	\draw (base.east) -- ($(base.east)+(5mm,0)$);
	%\node[align=center,left=2mm of init] (inittext) {Initial state\\goes here};
	
	\draw ($(base.north)-(15mm,0)$) -- ($(base.north)+(-15mm,5mm)$);
	
	\draw ($(base.north)+(15mm,0)$) -- ($(base.north)+(15mm,5mm)$);
	
	\draw ($(base.north)+(-15mm,10mm)$) -- ($(base.north)+(-15mm,15mm)$) node[box,above] (env1) {MPO for\\
	environment 1} -- ($(env1.north)+(0,5mm)$);
	
	\draw ($(base.north)+(15mm,10mm)$) -- ($(base.north)+(15mm,15mm)$) node[box,above] (env2) {MPO for\\
	environment 2} -- ($(env2.north)+(0,5mm)$);
	
	\end{tikzpicture}.
	\end{equation}
	Each MPO is that computed by contracting the lower $k$ rows in Eq.~\eqref{network:fulllegsmoved}. The dynamical map to time $t_k$ is found by contracting these MPOs into the rank-4 tensor and then contracting trace caps into all of their remaining legs.

	\subsection{Partial coarse graining}
	There are three convergence parameters associated with each process tensor: the discretisation timestep $\Delta$, the singular value threshold $\chi$ for truncation and the memory length $K$ which dictates how many timesteps of system history are kept to capture the non-Markovian influence. In practice this final approximation corresponds to setting $\mathcal{I}_{k>K}=\mathds{1}\otimes \mathds{1}$. For the two baths considered in this paper the timescales of the different bath correlation functions are two orders of magnitude apart. Na\"ively the timestep appears to be fixed by the fastest bath (the optical bath in this case) and on this timescale there is no possibility of making a memory cutoff for the slower (vibrational) bath.
	
\begin{figure*}
		\centering
		\includegraphics[width=0.8\linewidth]{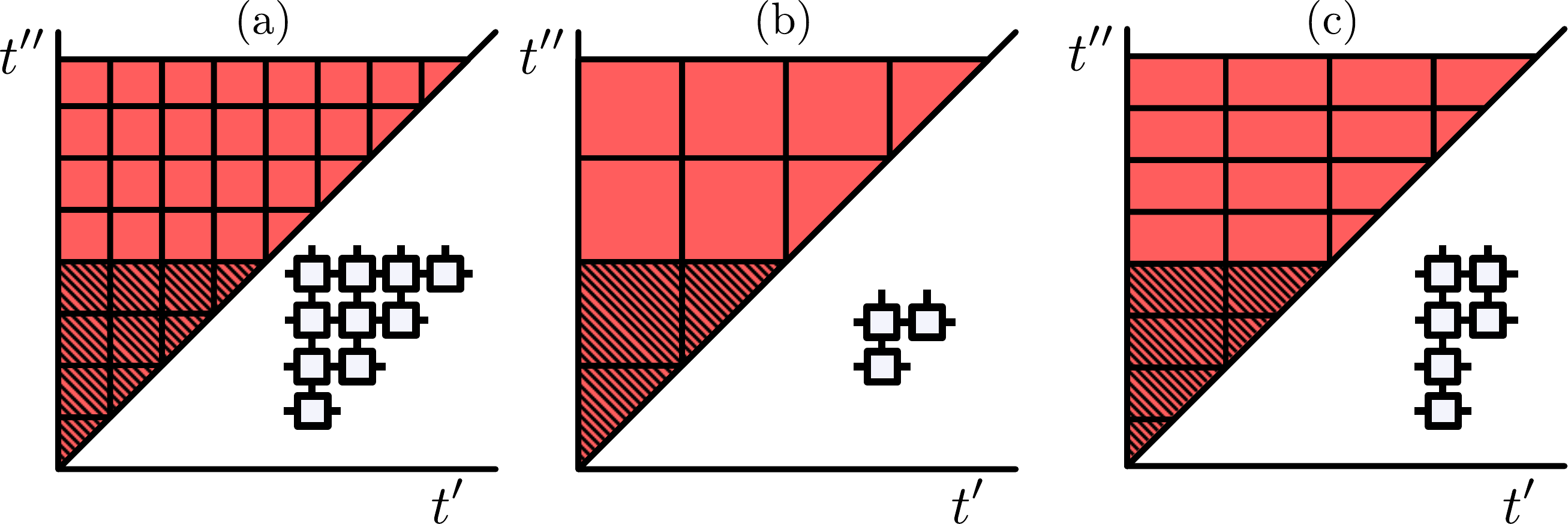}
		\caption{Visualization showing the discretization of the double integral in Eq.~\eqref{eq:eom} along with networks corresponding to the hatched areas. We have (a) the same timestep for both integrals, (b) same as (a) but with a timestep twice as large and (c) the inner integral with a timestep twice that of the outer integral.}
		\label{fig:new_time_disc}
	\end{figure*}

	Considering how the separate process tensors are contracted it might be thought that a larger timestep could be used for the slower bath with the same time-ordering contractions applied, i.e.~with the faster bath being internally contracted, as in \eqref{eq:dynmapcalc2}, several times between contractions with the slower bath. We found that while this approach converges to the correct result the error grows too quickly with increasing disparity in timesteps to be useful. We did, however, find a solution that enabled us to propagate to much later times. Comparing Figures \ref{fig:new_time_disc}(a) and (b) we illustrate how doubling the timestep manifests in both the discretisation of Eq.~\eqref{eq:discretize} and the resulting tensor network. Figure \ref{fig:new_time_disc}(c) shows how we can discretise the inner and outer integrals of Eq.~\eqref{eq:discretize} with different timesteps. We will refer to the timesteps of the inner and outer integrals as the history $\Delta_{h}$ and propagation $\Delta_{p}$ timesteps respectively. For certain parameter regimes we found that we achieved convergence with factors  $\Delta_{h}/\Delta_{p}$ on the order of 100 which allowed propagation lengths on the order of 1000 with no memory cutoff for the vibrational environment.
	
	\section{Incoherent population transfer}\label{Sec:modelandmethods}
	\subsection{Hamiltonian}\label{Sec:model}
	As an example application of multi-bath TEMPO, we consider a single two-level dipole that is coupled to both an optical and vibrational environment. 
% 	Competition between the two environments leads to rich and complex nonequilibrium behaviour that is ubiquitous to condensed matter systems. 
% 	A key example, is the physics of light-harvesting systems where surrounding protein structure leads to vibrational effects \cite{hedley2017light,adronov2000light, felip2016chameleonic, burzuri2016sequential,scholes2011lessons}. 
	It has recently been shown that the two baths in this system are non-additive, and when treated rigorously can lead to population inversion of the two-level dipole \cite{maguire2019environmental}. 
	Therefore, this is the ideal proving ground for testing our approach.
	%The optical interaction leads to transitions between the electronic states of the dipole by formation or annihilation of an electron-hole pair (exciton), whilst the vibrational coupling causes the manifolds to become displaced relative to each other. 
	
	The Hamiltonian describing this system can be partitioned into system ($S$), vibrational ($V$) and optical ($O$) parts as $H=H_S+H_V+H_O$. The system part,
	\begin{align}\label{eq:HS}
	H_S=\delta\sigma^+\sigma^-,
	\end{align}
	describes the electronic states of the dipole with transition energy $\delta$. The vibrational part has two contributions:
	\begin{equation}\label{eq:HV}
	H_V=\sum_{\mb{k}}\omega\uk b\uk^{\dagger}b\uk+\sigma^+\sigma^-\sum_{\mb{k}}\left(g\uk b\uk^{\dagger}+g\uk^*b\uk\right),
	\end{equation}
	where the first term describes the energy of the bath alone and the second is the electron-phonon interaction term. 
    The vibrational bath is composed of phonons of wavevector $\mb{k}$ and energy $\omega\uk$, with ladder operators $b\uk$ and $b\uk^{\dagger}$. 
	Each of these modes couples to the dipole with strength $g\uk$ as described by the interaction term. The effect of the bath on the system is fully characterised by the phonon spectral density $J_V(\omega)=\sum\uk\left|g\uk\right|^2\delta(\omega-\omega\uk)$, the exact functional form of which will be considered in later sections. 
	This interaction induces a displacement of the excited electronic manifold of the dipole which leads to Franck-Condon physics, that is, the non-trivial overlap of the vibrational wavefunctions between the manifolds \cite{kok2010introduction,maguire2019environmental}. %In Eqn.~\eqref{eq:HV} we do not fix the number of modes $N$. In Section~\ref{Sec:polaron} we look at the case of a single mode ($N=1$) and afterwards focus on the more physically realistic case of a continuum of modes ($N=\infty$).
	
	Finally, within the electric dipole approximation, the optical contribution to the Hamiltonian is given by:
	\begin{equation}
	H_O=\sum\uq\nu\uq a\uq^{\dagger}a\uq+\sigma^x\sum\uq \left(f\uq a\uq^{\dagger}+f\uq^* a\uq\right),\label{eq:HO}
	\end{equation}
    where we have introduced the photon ladder operators $a\uq$ and $a\uq^{\dagger}$, corresponding to photons with energy $\nu\uq$ and wavevector $\mb{q}$.
	This interaction term describes absorption or emission of a photon along with creation or annihilation of an exciton in the dipole. The light matter coupling strength $f\uq$ leads to the optical spectral density $J_O(\nu)=\sum\uq\left|f\uq\right|^2\delta(\nu-\nu\uq)$.
	The form of the optical spectral density is dependent on the gauge chosen for the light-matter interaction \cite{stokes2012extending,stokes2018master}. 
	As we will see, whether or not it is possible for this model to exhibit population inversion is determined by the form of $J_O$. 
	In this paper we will use 
	\begin{equation}
	J_O (\nu) = \alpha \frac{\nu^p}{\nu_c^{p-1}} e^{-\nu/\nu_c},
	\label{eq:JO}
	\end{equation}
	where $\alpha>0$ is the light matter coupling strength constant. We have also introduced a phenomenological cut-off $\nu_c$ which is required for the influence functional in TEMPO to converge, and is also physically justified for finite systems~\cite{stokes2018master, drummond1987unifying}.
	The Ohmicity of the spectral density $p$ is determined by the gauge choice and takes a positive value. In this work, we shall consider the two values of $p$ corresponding to two gauges that are most common to non-relativistic quantum electrodynamics, the $p=1$ (Coulomb) and ($p=3$) (multipolar) gauges \cite{stokes2012extending,stokes2018master}.

	\subsection{Population inversion mechanism}\label{Sec:simple}
\begin{figure}[hb!]\centering
	\includegraphics[width=0.4\textwidth]{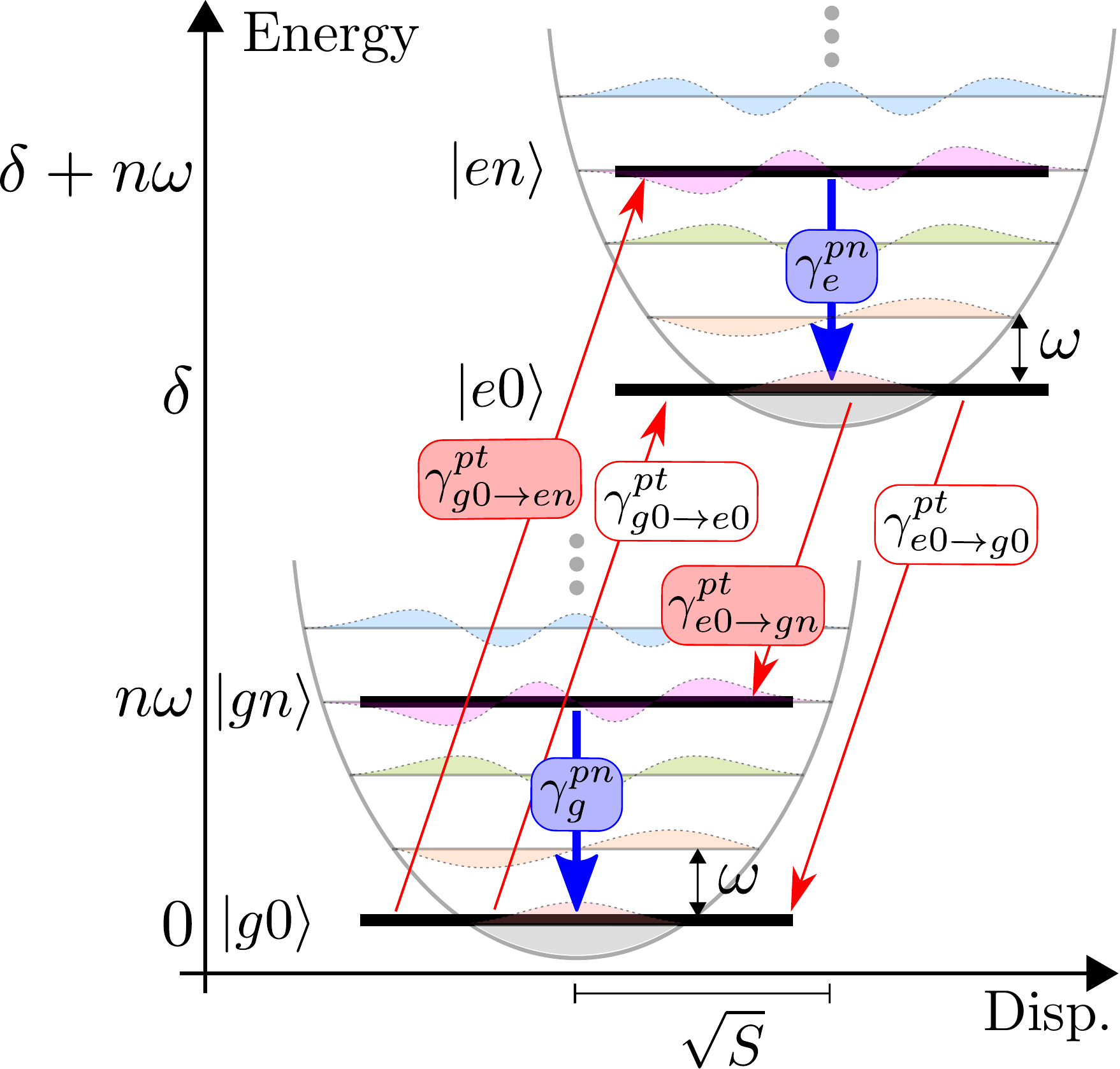}
		\caption{The simple four level model used to describe the mechanism of population inversion. Red and blue arrows denote photon and phonon transitions respectively. The arrows with coloured labels identify the pathway that creates population inversion. The symbols are defined in the main text.} \label{fig:simplemodel}
	\end{figure}

	To give an intuitive understanding of the population inversion mechanism and demonstrate its reliance on non-additivity of the two environments we first consider a simplified four level model shown in Fig.~\ref{fig:simplemodel}.
	This model consists of two electronic states separated by energy $\delta$, interacting with a single vibrational mode with frequency $\omega$ and coupling strength $g$.
	We then make three key approximations: we first assume that only two vibrational states contribute to the dynamics of the system, the zeroth and $n^{\text{th}}$ levels; second, the vibrational environment is assumed to be at zero temperature; and third, vibrational processes occur on a timescale much faster than optical transitions, such that photon emission and absorption only occur from the $\ket{e0}$ and $\ket{g0}$ states, as shown in Fig.~\ref{fig:simplemodel}.
% 	in comparison to the full system in Section~\ref{Sec:model}, we assume that there is one dominate phonon mode with energy $\omega$ and coupling $g$. 
%The impact of the vibrational mode on the system is quantified by the dimensionless Huang-Rhys parameter $S=(\left|g\right|/\omega)^2$. 
%Additionally, we assume that in each manifold the vibrational coupling only induces transitions between the ground and one excited vibrational level, which has an energy $n\omega$ greater than the ground. 
% Two further simplifications are made. First that the vibrational mode is at zero temperature so that vibrational processes cannot cause gains in energy. Second, that vibrational processes are much quicker than optical ones, in which case optical transitions will only occur from the $\ket{g0}$ and $\ket{e0}$ states. This condition is often fulfilled \cite{scerri2017method,nazir2016modelling}.
	
	Writing down a simple rate equation for the populations of the four level system, and solving for the steady states, we find that the ratio of the steady state populations of the electronic states are
	\begin{equation}\label{eq:simplesteady}
	P=\frac{\gamma^{pt}_{g0\to e0}+\gamma^{pt}_{g0\to en}}{\gamma^{pt}_{e0\to g0}+\gamma^{pt}_{e0\to gn}}.
	\end{equation}
	For population inversion we require that $P>1$. Using Fermi's Golden Rule, we find the rates
	\begin{subequations}\label{eq:simplemodelrates}
		\begin{align}
		&\gamma^{pt}_{g0\to e0}=2\pi W_0J_A(\delta),\\
		&\gamma^{pt}_{g0\to en}=2\pi W_nJ_A(\delta+n\omega),\\
		&\gamma^{pt}_{e0\to g0}=2\pi W_0J_E(\delta),\\
		&\gamma^{pt}_{e0\to gn}=2\pi W_nJ_E(\delta-n\omega),
		\end{align}
	\end{subequations} where $W_{j}=S^j\exp(-S)/j!$ are the Franck-Condon factors accounting for the overlap of the displaced vibrational levels between manifolds differing in number by $j=0,n$, and $S=(\left|g\right|/\omega)^2$ is the dimensionless Huang-Rhys parameter. 
	We have also introduced the emission and absorption optical spectral densities
 	\begin{subequations}\label{eq:JAJE}
		\begin{align}
		&J_E(\nu)=J_O(\nu)\left[1+N_O(\nu)\right],\label{eq:JE}\\
		&J_A(\nu)=J_O(\nu)N_O(\nu),\label{eq:JA}
		\end{align} 
	\end{subequations}
	where $N_O(\nu)=[e^{-\nu/T_O}-1]^{-1}$ is the population of the photon mode with energy $\nu$ at the temperature of the optical bath $T_O$. 
	
	When the vibrational coupling is weak ($S\approx 0$) the manifolds are displaced only slightly such that $W_0\gg W_n$. In this limit
	\begin{equation} 
	\lim_{W_0\gg W_n}P=\frac{N_O(\delta)}{[1+N_O(\delta)]},
	\end{equation} 
	where we have inserted Eq.~\eqref{eq:JO} for $J_O$. This approaches $1$ from below as $T_O\to\infty$ and so population inversion does not occur. Clearly then, strong vibrational coupling is key to population inversion and in this limit we instead find
	\begin{equation}\label{eq:SS}
	\lim_{W_n\gg W_0}P\approx\frac{N_O(\delta+n\omega)(\delta+n\omega)^p}{[1+N_O(\delta-n\omega)](\delta-n\omega)^p}.
	\end{equation}
	Since $W_n$ is maximised for $n=S$ (here approximating $n$ as continuous)
	we should replace $n$ with $S$ in Eq.~\eqref{eq:SS}. 
	From Eq.~\eqref{eq:SS}, we then see that the reason we need strong vibrational coupling is first to suppress optical transitions $\propto W_0$, i.e. to enhance the transition pathway identified by coloured labels in Fig.~\ref{fig:simplemodel}, and second to increase the asymmetry in the transition energy of the decay and excitation rates along this pathway. This asymmetry can be enhanced in two additional ways: increasing either $T_O$ or $p$ to make $J_E$ and $J_A$ more sensitive to energy changes. Since population inversion is most easily achieved at high optical temperatures when $N_O\gg1$, we focus on this limit and find
	\begin{equation}\label{eq:PStrongHigh}
	\lim_{\substack{W_n\gg W_0\\T_O\to\infty}}P\to\left(\frac{\delta+S\omega}{\delta-S\omega}\right)^{p-1}.
	\end{equation}
	Here we see one very important prediction of the model: no matter how strong the vibrational coupling, population inversion cannot occur for $p\le1$. Since this is true at infinite optical temperature when inversion is easiest, this must be true for all temperatures. This result will be emphasised when we compare multi-bath TEMPO to the reaction coordinate and polaron calculations, introduced in the next section, across all parameter regimes in Section~\ref{Sec:Comparisons}. %This leads to a qualitative gauge ambiguity when comparing the Coulomb ($p=1$) and multipolar ($p=3$) gauges for this model.
	
	Eq.~\eqref{eq:PStrongHigh} is independent of the optical coupling strength. There are two reasons for this. First, the Fermi Golden rule rates are only valid for small optical coupling. Second, we have assumed that the vibrational rates are significantly faster than the optical rates. In the next section, we will derive the master equations for the polaron and RC techniques. The polaron technique similarly treats the optical coupling to second order and assumes that the vibrational bath is displaced instantaneously. This is the same as assuming that vibrational rates are large enough to lead to effectively instantaneous transitions. The RC technique relies on the weak light-matter coupling but does not assume that the vibrational bath displaces instantaneously, and in Section~\ref{Sec:Comparisons} we find that population inversion decreases as optical lifetime becomes comparable to the displacement timescale.

	\section{Approximate techniques}
	\label{Sec:approxtech}
	In this section we will give a brief introduction to the two approximate techniques, a master equation that exploits the polaron transformation and the RC mapping. Both approaches are capable of capturing non-perturbative electron-phonon effects, however, they operate in two distinct regimes as will be discussed below.% in the following section. 
	%\jake{Maybe: Provide insight into two distinct but complementary forms of vibrational environment.} A significant difference between the two techniques is for which vibrational spectral densities they work well for.

	\subsection{Polaron theory}
	\label{Sec:polaron}
% 	For the system described in Section~\ref{Sec:model}, the polaron transformation works well for any vibrational spectral density whose weighted moments, 
% 	\begin{equation}\label{eq:weightedmo}
% 	\mu_j=\inta{\omega}\frac{J_V(\omega)}{\omega^2}\omega^j,
% 	\end{equation}
% 	are finite and $\mu_j/j!$ decays sufficiently quickly as $j$ increases. This includes the widely used class of spectral densities with general form
% 	\begin{equation}\label{eq:JVPol}
% 	J_V(\omega)\propto\omega^a\exp\left[-(\omega/\omega_c)^b\right],
% 	\end{equation}
% 	for $(a,b)>(1,0)$ if $\omega_c$ is small. These are typical to the environments of quantum dots \cite{nazir2016modelling}.

	Polaron theory accounts for strong electron-phonon interactions through a unitary transformation to the global Hamiltonian, such that $H_P=UHU^{\dagger}$ where $U=\mathrm{exp}\left[-G\sigma^+\sigma^-\right]$ with
	\begin{equation}\label{eq:poldisplaceop}
	G=\sum\uk\left[\frac{g\uk}{\omega\uk}b\uk^{\dagger}-\frac{g\uk^*}{\omega\uk}b\uk\right].
	\end{equation}
	This transformation dresses the excitonic degrees of freedom with vibrational modes of the environment forming a polaron, and providing a transformed basis in which to do perturbation theory. This has the advantage of providing a simple analytic expression for the master equation, while still accounting for strong system-environment interactions~\cite{mccutcheon2010quantum,nazir2016modelling}.
% 	The aim of this technique is to remove the vibrational interaction by moving to the polaron frame. Here, the two electronic states of the dipole no longer describe whether or not an exciton is present. They now refer to polarons; quasiparticles made up of the exciton dressed by the vibrational interaction energy. The population dynamics of the polaron and exciton are the same which means that we only need to track the polaron. This has the advantage that a second order polaron master equation will not break down at strong vibrational coupling because the energy associated with the vibrational interaction is no longer treated perturbatively.
	%
% 	The polaron frame Hamiltonian is obtained by the unitary transformation $H_P=UHU^{\dagger}$ where $U=\mathrm{exp}\left[-G\sigma^+\sigma^-\right]$ with
% 	\begin{equation}
% 	G=\sum\uk\left[\frac{g\uk}{\omega\uk}b\uk^{\dagger}-\left(\frac{g\uk}{\omega\uk}\right)^*b\uk\right].
% 	\end{equation}
	The resulting Hamiltonian in the polaron frame reads:
	\begin{align}\label{eq:HP}
	H_{P}=&\delta_{P}\sigma^+\sigma^-+\sum\uk\omega\uk b\uk^{\dagger}b\uk+\sum\uq\nu\uq a\uq^{\dagger}a\uq\nonumber\\
	&+\left(B^{\dagger}\sigma^++B\sigma^-\right)\sum\uq \left(f\uq a^\dagger\uq+f^*\uq a\uq\right),
	\end{align}
% 	where the polaron energy $\delta_{P}=\delta-\lambda$ with reorganisation energy $\lambda=\inta{\omega}J_V(\omega)/\omega$ and we have defined the displacement operator $B=\mathrm{exp}\left[-G\right]$. The mathematical details of this transformation are well documented and can, for example, be found in references \cite{kok2010introduction,brandes2005coherent,nazir2009correlation,qin2017effects,nazir2016modelling,pollock2013multi,rouse2019optimal}.
	%
	where the transformation has removed the linear vibrational coupling. The dipole transition operators are now dressed with the displacement operators $B=\mathrm{exp}\left[-G\right]$, such that $B^\dagger\sigma^+$ ($B\sigma^-$) creates (destroys) a polaron with energy $\delta_P=\delta-\lambda$, where $\lambda=\inta{\omega}J_V(\omega)/\omega$ is the reorganisation energy. The mathematical details of this transformation are well documented and can, for example, be found in references \cite{mccutcheon2010quantum,roy2011phonon,roy2012polaron,wilson2002quantum,kok2010introduction,brandes2005coherent,nazir2009correlation,qin2017effects,nazir2016modelling,pollock2013multi,rouse2019optimal}. 
	%the exciton energy is replaced with the polaron energy, and the original $\sigma^x$ coupling to the dipole in the optical interaction has gained factors of the displacement operator. This accounts for the upper electronic manifold having been displaced by the vibrational interaction compared to the lower manifold; optical transitions either start or end in the displaced manifold, creating ($B^{\dagger}\sigma^+$) or annihilating ($B\sigma^-$) polarons.

	To describe the dynamics generated from Eq.~\eqref{eq:HP} we derive a Born-Markov master equation in which the optical and vibrational degrees of freedom are captured to second order in the polaron frame. 
% 	for polaron formation (and equivalently exciton formation) we can derive the second order Born-Markov master equation in the usual way \cite{breuer2002theory}. 
	In this particular example, the populations are independent of the coherences of the system, leading to a simple master equation of the form: 
	%doing so, one finds that the excited and ground state populations, $\rho_{ee}$ and $\rho_{gg}$, are described by the master equations
	\begin{equation}\label{eq:rhoeePol}
	\dot{\rho}_{ee}(t)=\gamma_{\uparrow}\rho_{gg}(t)-\gamma_{\downarrow}\rho_{ee}(t),
	\end{equation}
	where $\rho_{ee}$ and $\rho_{gg}$ are the excited and ground state populations in the lab frame.
% 	and $\dot{\rho}_{gg}(t)=-\dot{\rho}_{ee}(t)$. We have not made any secular approximation; the populations are naturally decoupled from the coherences for this system. 
	We have also defined the excitation $\gamma_{\uparrow}=\gamma(-\delta_{P})$ and decay $\gamma_{\downarrow}=\gamma(\delta_{P})$ rates 
% 	are defined as
% 	\begin{subequations}\label{eq:garrow}
% 		\begin{align}
% 		&\gamma_{\uparrow}=\gamma(-\delta_{P}),\label{eq:gexcite}\\
% 		&\gamma_{\downarrow}=\gamma(\delta_{P}),\label{eq:gdecay}
% 		\end{align}
% 	\end{subequations}
 		in terms of the polaron rate function
	\begin{equation}\label{eq:ECF}
	\gamma(\zeta)=2\mathrm{Re}\inta{t}e^{i\zeta t}\mathcal{B}(t)\mathcal{C}(t),
	\end{equation}
 where the phonon correlation function is
	\begin{equation}
	\mathcal{B}(t) = \mathrm{Tr}_V\left[B^{\dagger}(t)B(0)\rho_V\right]=\rme^{-\phi(0)}\rme^{\phi(t)},
	\end{equation}
	defined in terms of the Gibbs state of the vibrational environment, $\rho_V$, and the displacement operator in the interaction picture, $B(t)$. The phonon correlation function is characterised by the phonon propagator
	\begin{align}\label{eq:phi}
	&\phi(t)=\nonumber\\
	&\int\limits_0^\infty \text{d}\omega\  \frac{J_V(\omega)}{\omega^2}\left[\coth\left(\frac{\omega}{2 T_V}\right)\cos(\omega t) - i\sin(\omega t) \right],
	%\sum\uk S\uk\bigg[\cos\left(\omega\uk t\right)&\coth\left(\frac{\beta_V\omega\uk}{2}\right)\nonumber\\
	%&\hspace{1.5cm}-i\sin\left(\omega\uk t\right)\bigg],
	\end{align}
	where $T_V$ is the temperature of the vibrational bath and we have taken a continuum limit for the phonon modes \cite{mccutcheon2010quantum,roy2011phonon,roy2012polaron,kok2010introduction,brandes2005coherent,nazir2009correlation,qin2017effects,nazir2016modelling,pollock2013multi,rouse2019optimal}. When comparing TEMPO with polaron theory, we shall take the spectral density
	\begin{equation}
	J_V (\omega) = S \frac{\omega^3}{\omega_c^2} e^{-\omega/\omega_c},
	\label{eq:JV_pol}
	\end{equation}
	where $S$ is the dimensionless Huang-Rhys parameter and $\omega_c$ is the phonon cut-off frequency.% The super-Ohmic form of the spectral density is necessary to remove the infra-red divergence observed in polaron theory.

	The optical contribution to the rate is given by the correlation function
	\begin{align}\label{eq:CC}
	\mathcal{C}(t) &=\sum_{\mb{q}\mb{q}^\prime}\text{Tr}_O\left[A\uq^\dagger(t)A\uqp(0)\rho_O\right]\\
	%= \sum\limit\uq\mathrm{Tr}_O\left[C\uq(t) C\uq\rho_O\right]=
	&=\sum\uq \vert f\uq \vert^2\left(N_O(\nu\uq)\rme^{i\nu\uq t}+\left[N_O(\nu\uq)+1\right]\rme^{-i\nu\uq t}\right),\nonumber
	\end{align}
    where $\rho_O$ is the Gibbs state of the optical environment and $A\uq(t)$ is the operator $f\uq a^\dagger\uq+f^*\uq a\uq$ in the interaction picture. Taking the continuum limit over the optical modes, we obtain an expression in terms of the absorption and emission effective spectral densities
	\begin{equation}\label{eq:ECF2}
	\gamma(\zeta)=2\inta{\nu}\Big[ J_E(\nu)\mathcal{V}(\zeta-\nu)+J_A(\nu)\mathcal{V}(\zeta+\nu)\Big],
	\end{equation}
	where the vibrational influence is contained within~\cite{mccutcheon2013model,RoyChoudhury2015Fermi,Roy2015emission}
	\begin{equation}\label{eq:K}
	\mathcal{V}(\epsilon)=\rme^{-\phi(0)}\mathrm{Re}\inta{t}\rme^{\phi(t)}\rme^{i\epsilon t}.
	\end{equation}
    The non-additive nature of the vibrational and optical interactions is now clear because Eq.~\eqref{eq:ECF2} is a convolution of optical functions with a vibrational function. Eq.~\eqref{eq:ECF2} can be solved numerically for the optical and vibrational spectral densities considered in this paper.
	
	 %To arrive at analytic results, it is  typical to assume that the optical spectral density is independent of frequency around the emission energy \cite{nazir2016modelling,scerri2017method}. This approximation results in $\mathcal{K}(\epsilon)\to\pi\delta(\epsilon)$ such that the vibrational contribution to the optical rates is simply to change the sampling energy from the electronic energy $\delta$ to the polaron energy $\delta_P$. In this case, the non-additive effects of vibrational coupling in the polaron frame are lost and population inversion becomes impossible~\cite{maguire2019environmental}. Instead, to capture the non-additivity, we derive for the first time a full analytic solution to Eqn.~\eqref{eq:ECF2}, presented in SI~3. 

	\subsection{Reaction coordinate}\label{Sec:CC}
	An alternative semi-analytic approach for describing the model outlined in Section~\ref{Sec:model} is the RC method. Here key environmental degrees of freedom are included in an enlarged system Hamiltonian by way of a normal mode mapping~\cite{garg1985effect,thoss2001hybrid,hughes2009effective,martinazzo2011universal, iles2014environmental}. The remaining environmental degrees of freedom are then captured through a residual environment, which can be treated to second order using a Born-Markov master equation~\cite{iles2014environmental,iles2016energy}. 
	This treatment allows one to access strong coupling and highly non-Markovian regimes with the conceptual simplicity of a master equation, and little computational cost~\cite{gribben2020exact}.
	In this section we shall give an outline of the RC mapping and the subsequent derivation of the master equation. For more details we refer the reader to~\cite{iles2014environmental}.
	
	We shall start by restricting the phonon environment to the case of an underdamped spectral density, of the form:
	\begin{equation}\label{eq:JVLorentz}
	J_{\text{V}}(\omega) = \frac{\alpha_{\text{UD}}\Gamma\omega_0^2 \omega}{(\omega_0^2-\omega^2)^2+\Gamma^2\omega^2},
	\end{equation}
	which is one of the spectral densities for which the RC mapping is exact~\cite{iles2016energy}.
	Here $\alpha_{\text{UD}}$ is the reorganisation energy, $\omega_0$ is the frequency of the spectral density's peak and $\Gamma$ is its width. 
	
	Using a collective coordinate mapping, the vibrational contribution to the Hamiltonian given in Eq.~\eqref{eq:HV} becomes $H_\mathrm{V}\mapsto H_\mathrm{RC} + H_\mathrm{Res}$, where
	\begin{align}
	H_\mathrm{RC} &=\eta \sigma^+\sigma^- (a^\dagger + a) + \Omega a^\dagger a,\\
	H_\mathrm{Res}&=(a^\dagger + a)\sum_\mathbf{k} h_\mathbf{k}\left(c_\mathbf{k}^\dagger + c_\mathbf{k}\right) 
	+ \sum_\mathbf{k} \nu_\mathbf{k} c^\dagger_\mathbf{k}c_\mathbf{k}\\	
	&\quad+(a^\dagger + a)^2 \sum_\mathbf{k}\frac{h_\mathbf{k}^2}{\nu_\mathbf{k}}. \nonumber
	\end{align}
	Here we have introduced the annihilation (creation) operators $a$ $(a^\dagger)$ and $c_\mathbf{k}$ ($c_\mathbf{k}^\dagger$) for the RC and residual environment respectively. The mapped parameters can be written in terms of the unmapped spectral density, where  $\eta = \sqrt{\pi\alpha_\mathrm{UD}\omega_0/2}$, $\Omega = \omega_0$, and the residual environment is described by the spectral density $J_\mathrm{Res}(\omega) =\sum_\mathbf{k} \vert h_\mathbf{k}\vert^2\delta(\omega - \nu_\mathbf{k})=\Gamma\omega/2\pi\Omega$.
	
	Since the collective coordinate mapping acts only on the modes of the vibrational environment, leaving the system and optical degrees of freedom unchanged, then in the mapped frame we have: $H = \tilde{H}_\mathrm{S} + H_\mathrm{Res} + H_\mathrm{O}$, where we have defined the augmented system Hamiltonian as $\tilde{H}_\mathrm{S} = H_\mathrm{S} + H_\mathrm{RC}$.
	
	A key advantage of the RC method is that we can treat the augmented system non-perturbatively, while treating the residual environment to second order, thus effectively capturing non-Markovian effects.
	To do this, we first apply the single mode polaron transformation to Hamiltonian $U = \exp[-\nu^{-1}\eta\sigma^{+}\sigma^{-}(a^\dagger - a)]$.
	This transformation diagonalizes the augmented system Hamiltonian such that $H^\prime_\mathrm{S} = U\tilde{H}_\mathrm{S}U^\dagger= \delta \sigma^+\sigma^- + \Omega a^\dagger a$.
	In this transformed frame,
% 	We do this using a Born-Markov master equation, which, following the derivation outlined in~\cite{maguire2019environmental}, 
we obtain a master equation of the form $\partial_t \rho(t) = \mathcal{L}[\rho(t)]$, with the Liouvillian:
	\begin{equation}
	\mathcal{L}[\rho(t)] = -i\left[\tilde{H}^\prime_\mathrm{S},\rho(t)\right] + \mathcal{K}_\mathrm{Res}[\rho(t)] + \mathcal{K}_\mathrm{O}[\rho(t)],
	\end{equation}
	where $\rho(t)$ is the density operator for the augmented system.
	Here $\mathcal{K}_\mathrm{Res}$ is a superoperator representing the action of the residual environment~\cite{iles2014environmental}:
	\begin{equation}
	\mathcal{K}_\mathrm{Res}[\rho] = \left[\mathcal{A},\rho(t)\zeta\right] + \left[\zeta^\dagger\rho(t),\mathcal{A}\right],
	\end{equation}
	where $\mathcal{A} = a^\dagger + a -2(\eta/\nu)\sigma^+\sigma^-$, and we have introduced the rate operator:
	\begin{equation}
	\zeta = \frac{\pi}{2}\sum_{\lambda}J_\mathrm{Res}(\lambda)\left[\coth\left(\frac{\lambda}{2 T_V}\right)+1\right]\mathcal{A}(\lambda),
	\end{equation}
	where the sum runs over the eigensplittings of the augmented system Hamiltonian, such that $\lambda \in\{\pm\Omega, 0\}$, with the corresponding transition operators $\mathcal{A}(0) = -2(\eta/\nu)\sigma^+\sigma^-$, $\mathcal{A}(\Omega) = a$, and $\mathcal{A}(-\Omega) = a^\dagger$.% \jake{I need to double check this...}.
	
	%we have introduced the eigenbasis of the augmented system $\tilde{H}_\mathrm{S}\ket{\psi_j} = \psi_j\ket{\psi_j}$, with $\lambda_{jk} = \psi_j - \psi_k$, and $\mathcal{A}_{jk} = \bra{\psi_j}\mathcal{A}\ket{\psi_k}$.
	
	Crucially, unlike standard perturbative treatments, the RC is capable of accounting for non-additive effects when multiple environments are present~\cite{maguire2019environmental}. In the case of an optical environment, this is done by explicitly deriving the optical contribution to the master equation in the eigenbasis of the augmented system.
	This results in a dissipator of the form:
	\begin{equation}
	\mathcal{K}_\mathrm{O}[\rho(t)] = - \left[\sigma^x,\left(\chi_\uparrow+\chi_\downarrow\right)\rho(t)\right] +\mathrm{h.c.}, 
	\end{equation}
	where we have defined the rate operators corresponding to absorption $\chi_\uparrow = \sum_{jk}\sigma^x_{jk}\gamma_\mathrm{RC}^{\uparrow}(\lambda_{jk})\ket{\psi_j}\bra{\psi_k}$, and emission $\chi_\downarrow = \sum_{jk}\sigma^x_{jk}\gamma_\mathrm{RC}^{\downarrow}(\lambda_{jk})\ket{\psi_j}\bra{\psi_k}$.
	The rates for these processes are given as:
	\begin{align}
	\gamma_\mathrm{RC}^\uparrow(\lambda)& =\pi J_A(\lambda)\Theta(\lambda),\label{eq:rc_up}\\
	\gamma_\mathrm{RC}^\downarrow(\lambda) &=  \pi J_E(-\lambda)\Theta(-\lambda),\label{eq:rc_down}
	\end{align}
	where $\Theta(x)$ is the Heaviside step function. Here, the steady state is independent of the Lamb shifts, allowing them to be neglected in the calculations we will show.
	
	If we compare the optical transition rates calculated in the RC model (Eqs.~\eqref{eq:rc_up} and~\eqref{eq:rc_down}), and those found in polaron theory in Eq.~\eqref{eq:ECF2}, we see some striking similarities, with vibronic contributions entering directly into the expressions for the rates.
	However, a key difference between the RC and polaron approach is that the displacement of the phonon environment in polaron theory is static, which leads to a breakdown of the perturbative master equation when the system dynamics approach the correlation time of the phonon environment~\cite{mccutcheon2010quantum}. 
	In contrast, no such approximation is made in RC theory, and the displacement of the RC may vary throughout the system evolution, such that the method is valid in regimes of fast system dynamics.
	This will be significant in regimes where the system light-matter coupling dominates over the vibrational interaction.

	\section{Non-additive physics in all coupling regimes}\label{Sec:Comparisons}
	
	With the methods described above, we may now explore the effects of non-additivity on the nonequilibrium steady state of the TLE.
	To do this 
	we will separate the analysis into three regimes of optical coupling strength: weak, moderate and strong. Weak coupling is defined as the thermalization timescale being shorter than the optical lifetime, and moderate coupling the reverse of this. In the strong coupling regime, a second order truncation of the optical coupling breaks down and we use multi-bath TEMPO to explore this novel regime of light-matter coupling.
	In each regime we shall compare the steady state population behaviour predicted by the analytic models against multi-bath TEMPO. 
	Analysing the steady state exposes the most striking consequences of the non-additive effects arising from our model. Further it has the added advantage of circumventing the awkward problem of defining equivalent initial states for a fair comparison of the short time dynamics, however, of course the dynamics are also available in all three approaches.
	
\begin{figure*}[ht!]
		\centering
		\includegraphics[width=.9\linewidth]{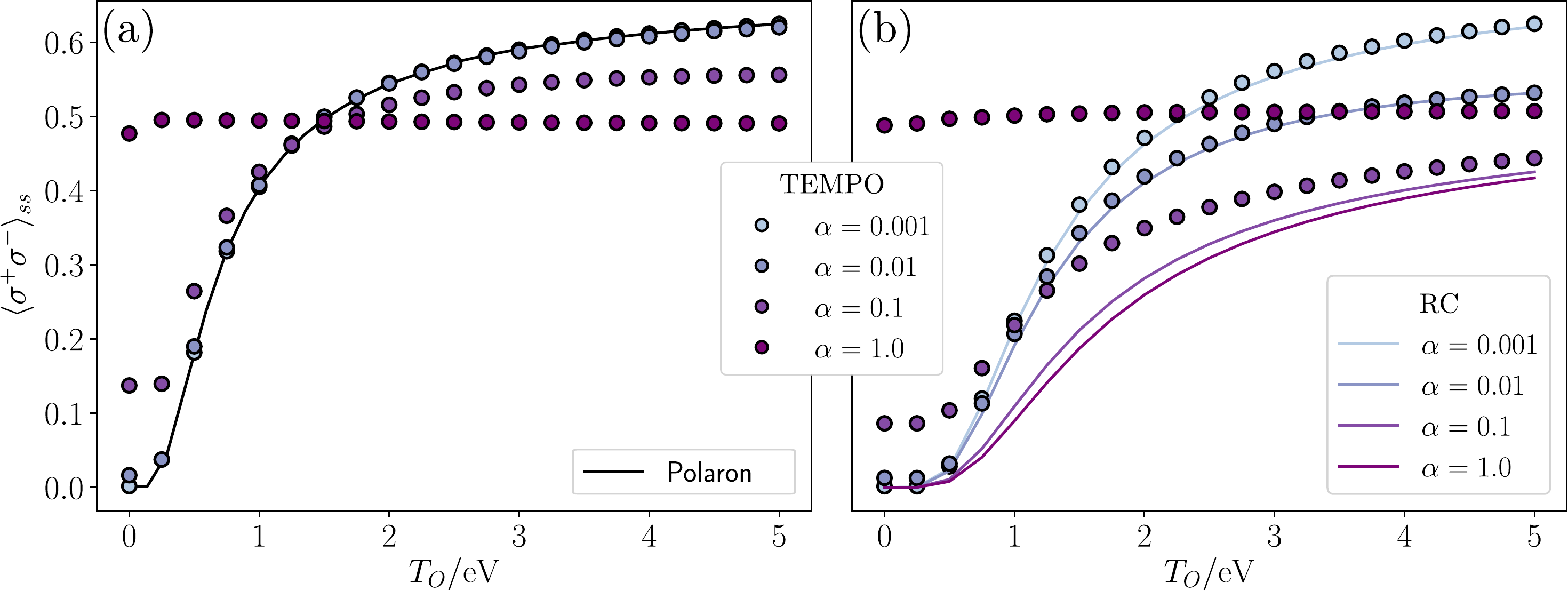}
		\caption{Steady state population as a function of the optical bath temperature $T_O$. In (a) we compare the populations as calculated by TEMPO and the polaron master equation for multiple optical couplings $\alpha$. The vibrational spectral density is of super-Ohmic form with an exponential cutoff as given by Eq.~\eqref{eq:JV_pol}. We fixed the Huang-Rhys factor at $S=1$, the cutoff to $\omega_c=0.101$~eV and the polaron splitting to $\delta_P=1$~eV. In (b) we plot the same comparison for the reaction coordinate (RC) master equation. The vibrational spectral density in this case is of Drude-Lorentz form as given by Eq.~\eqref{eq:JVLorentz}. We fixed the coupling constant at $\alpha_{\text{UD}}=0.3$~eV, the linewidth at $\Gamma=0.1$~eV, the mode frequency at $\omega_0=0.05$~eV and the polaron splitting at $\delta_P=2$~eV. The other parameters for both (a) and (b) are $p=3$, $\nu_c=10$~eV and $ T_V=0.0258$~eV.}
		\label{fig:polcomp}
	\end{figure*}

\begin{figure*}[ht!]
		\centering
		\includegraphics[width=.9\linewidth]{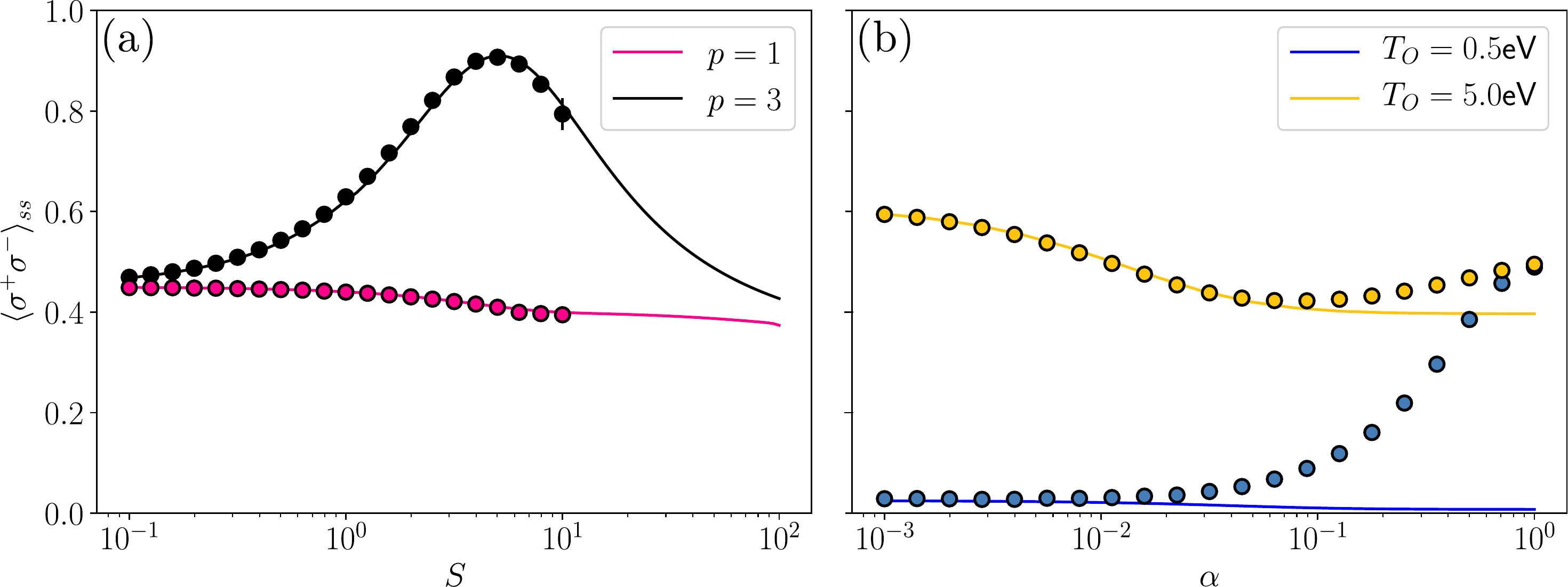}
		\caption{Steady state population as calculated using TEMPO (dots) and approximate techniques (lines). In (a) we compare TEMPO and the polaron master equation varying the Huang-Rhys factor for ohmicities $p=1$ and $p=3$ with fixed optical couplings, $\alpha=0.0001$ and $\alpha=0.001$ respectively, and temperature $ T_O= 5$~eV. The other parameters for (a) are $\delta_{P}=1$~eV and $\omega_c=0.101$~eV. In (b) we compare TEMPO and the RC master equation varying the optical coupling for two temperatures either side of the inversion transition. The other parameters for (b) are $\delta_{P}=2$~eV, $p=3$, $\alpha_{\text{UD}}=0.3$~eV, $\Gamma = 0.1$~eV and $\omega_0=0.05$~eV. The common parameters for (a) and (b) are $\nu_c=10$~eV and $ T_V=0.0258$~eV. }
		\label{fig:RCcomp}
	\end{figure*}

 	We present our comparisons for variable temperature of the optical field in Fig.~\ref{fig:polcomp} and as a function of light-matter coupling strength in Fig.~\ref{fig:RCcomp}. For all calculations, we use the optical spectral density defined in Eq.~\eqref{eq:JO}.
	For comparison with the polaron theory (Figs.~\ref{fig:polcomp}(a) and~\ref{fig:RCcomp}(a)), we use the super-Ohmic vibrational spectral density given in Eq.~\eqref{eq:JV_pol} as detailed in Sec.~\ref{Sec:polaron}, while Eq.~\eqref{eq:JVLorentz} is used for the RC comparisons in Figs.~\ref{fig:polcomp}(b) and~\ref{fig:RCcomp}(b).
	
% 	To simulate the optical environment with TEMPO we need to add a high frequency cutoff to the spectral density such that the integral to calculate the influence phase converges. As well as a necessity for computation this is also physically justified with a finite system. For all the results in this section we use a superohmic spectral density with an exponential cutoff:
% 	\begin{equation}
% 	J_O (\nu) = \alpha \frac{\nu^p}{\nu_c^{p-1}} e^{-\nu/\nu_c},
% 	\end{equation}
% 	where $\alpha$ is a dimensionless coupling constant we will use to set the optical coupling strength, $\nu_c$ is the cutoff frequency and $p$, the Ohmicity, determines the low frequency behaviour.
	
% 	 we compare the steady state populations as predicted by the polaron transform approach and TEMPO, shown in Fig.~\ref{fig:polcomp}. 
	 
% 	of the form:
% 	\begin{equation}
% 	J_V (\omega) = S \frac{\omega^3}{\omega_c^2} e^{-\omega/\omega_c}.
% 	\end{equation}
% 	where $S$ is the Huang-Rhys parameter~\jake{S shouldn't be dimensionless here, right?}, and $\omega_c$ is the phonon cut-off frequency.

% 	In Figure~\ref{fig:RCcomp} we present a similar comparison between TEMPO and the RC technique looking again at steady state populations. Here we use the spectral density in Eq.~\eqref{eq:JVLorentz}.
	
	\subsection{Weak optical coupling}
	For small light-matter coupling strength, we see excellent agreement between both approximate methods and TEMPO, as is apparent from Fig.~\ref{fig:polcomp}. 
	This is expected as both theories can non-perturbatively account for electron-phonon interactions, and the second order treatment of the optical coupling is sufficient for small $\alpha$.
	Most striking from these figures is that for small $\alpha$ all approaches predict a population inversion at large optical temperatures.
    This is as we expect from the work of Maguire~\emph{et al}~\cite{maguire2019environmental}, as well as the four level analysis in Sec.~\ref{Sec:simple}, where vibrational states induce an asymmetry in the optical rates, suppressing optical decay while enhancing the converse absorption process.  %allow the electron population to overcome the
%	 In Figures~\ref{fig:polcomp} and~\ref{fig:RCcomp} we see good agreement of the approximate techniques with TEMPO at small $\alpha$. 
	 
% 	 In Figure~\ref{fig:polcomp}a we see good agreement between TEMPO and polaron over the range of vibrational couplings. We explore this for both $p=1$ and $p=3$.
	 
	 Using the polaron theory, we can also see that this behaviour persists across a full range of vibrational couplings, agreeing well with TEMPO in Fig.~\ref{fig:RCcomp}(a).  This is with the exception of the very largest couplings where there were difficulties with getting converged results from TEMPO. As well as the inherent difficulties associated with simulating stronger couplings, there is also a significant drop in the population transfer rates meaning that longer propagation times are necessary to reach steady state. This suppression with strong vibrational coupling occurs because the system tries to excite into high lying vibrational states, requiring high energy photon modes to be thermally populated to complete the transition. We have included error bars that correspond to the change in our steady state prediction if we end the propagation $40$ timesteps earlier to give an idea of how much we would expect the prediction to vary if we were to propagate out further. Taking these errors into account, this suggests that polaron theory predicts accurate steady state populations over the full range of Huang-Rhys parameters. 
	  
	  We may also address the important prediction from the four level model in Sec.~\ref{Sec:simple} that population inversion is conditional on the Ohmicity of the optical field.  This is confirmed by polaron theory and TEMPO in Fig.~\ref{fig:RCcomp}(a), where we see that population inversion never occurs when $p=1$, but does for $p=3$. Here we also found that a smaller value of $\alpha$ was required for $p=1$ in order to match the weak-coupling result from polaron theory.
% 	  Importantly, here we see that even in the numerically exact calculation, population inversion never occurs for $p=1$ but does for $p=3$; a result predicted by the simple model in Section~\ref{Sec:simple}. 
	  %Our full model with $p=1$ and $p=3$ refer to the widely used Coulomb (``$\mb{p}\cdot\mb{A}$'') and multipolar (``$\mb{d}\cdot\mb{E}$'') gauges respectively. It is now well-documented that truncating the dipole to a finite (often two) energy states creates ambiguities in observables calculated in different gauges; however, this type of ambiguity is resolved when the dipole is allowed more energy levels \cite{stokes2019gauge,de2018breakdown,de2018cavity}. The population inversion ambiguity seen in Figure~\ref{fig:polcomp} will not be resolved by allowing more dipole energy states because it depends only on the structure of the optical spectral density. Since this ambiguity only arises under the non-additive nature of the two-bath interaction, this raises the important question of the gauge-dependence of the vibrational interaction.~\jake{Is this a Gauge ambiguity? Or just not using consistent gauges when deriving the Hamiltonian for the electron-phonon interaction?}
	 
	 \subsection{Moderate optical coupling}
	 At moderate optical coupling, where the optical timescales approach those of the phonon processes, we begin to see the population inversion predicted by  TEMPO reduce. We attribute this to a \emph{dynamical undressing} of the optical transition, where an emission and absorption of a photon occurs faster than  polarons can be created or destroyed during transitions between the ground and excited state. 
	 
	 This is most apparent when comparing the polaron and TEMPO calculations in Fig.~\ref{fig:polcomp}(a), where the steady-state population predicted by polaron theory is independent of the light-matter coupling strength $\alpha$. 
	 This can be seen from the polaron frame master equation in Eq.~\eqref{eq:rhoeePol}, where the steady state population is dependent only on the ratio of $\gamma_{\uparrow}/\gamma_{\downarrow}$, which has no $\alpha$ dependence.
	 Physically, this is a consequence of the static state dependent displacement (Eq.~\eqref{eq:poldisplaceop}) applied to the phonon environment, regardless of the system dynamics.
	 This and the approximations used to derive the master equation in Eq.~\eqref{eq:rhoeePol}, mean that within this theory it is always the case that a polaron is formed during an optical transition, regardless of how quickly this transition occurs. This results in the steady state population being completely insensitive to the light-matter coupling strength. 
	 This is a common issue for polaron-type master equations, and may be overcome by modifying the displacement through a variational treatment~\cite{mccutcheon2011variational,bundgaardnielsen2021nonmarkovian}. However, this is not possible when the two baths are out of equilibrium with each other.
	 
	 In contrast, since the RC is included non-perturbatively in the augmented system, its state may respond to the increased optical transition rate, such that in regimes that polaron formation is not energetically favourable the TLE remains undressed by phonon modes.
	 This flexibility in the theory means that RC continues to be in good agreement with TEMPO for moderate optical coupling strengths ($\alpha=0.01$), as seen in Fig.~\ref{fig:polcomp}(b).

	 \subsection{Strong optical coupling}
	 
	 For large light-matter coupling strengths the second-order treatment use to derive the  interaction with the optical field in both approximate techniques breaks down. As can be seen in Fig.~\ref{fig:RCcomp}(b), this leads to significant deviation between the RC method and multi-bath TEMPO, where at large $\alpha$  TEMPO predicts a trend towards $\langle \sigma^+\sigma^-\rangle = 0.5$ independent of optical temperature $T_O$. 
	 
	 In the strong coupling limit a system-environment interaction can be thought of as a continuous measurement process where the system is consistently projected onto the eigenbasis of the coupling operator; the ultra-strong coupling limit ($\alpha\rightarrow\infty$) can therefore be described using the quantum Zeno effect (QZE)~\cite{facchi2008quantum,cresser2021weak}. In this limit the total system plus environments evolve under an effective Hamiltonian given by:
	 \begin{equation}
	     H_{QZE}=\sum_i P_i H P_i,
	 \end{equation}
	 where $P_i$ is a projection onto the $i$th eigenstate of the optical interaction. Explicitly this is defined as:
	 \begin{equation}
	     P_i=\dyad{\phi_i}{\phi_i},
	 \end{equation}
	 where
	 \begin{equation}
	     \sigma^x\sum\uq \left(f\uq a\uq^{\dagger}+f\uq^* a\uq\right)\ket{\phi_i}=\phi_i \ket{\phi_i}.
	 \end{equation}
	 These eigenstates can be partitioned into system and bath components and written as \begin{equation}
	 \{\ket{\phi_i}\}=\{\ket{+}\otimes\ket{x_i},\ket{-}\otimes\ket{x_i}\},
	 \end{equation} where $\{\ket{+},\ket{-}\}$ are the eigenstates of $\sigma^x$ and $\ket{x_i}$ the eigenstates of $\sum\uq \left(f\uq a\uq^{\dagger}+f\uq^* a\uq\right)$.
	 
	 However given that none of the other parts of the Hamiltonian act on the optical degrees of freedom we only need to consider how the system component of $P_i$ affects the system operators in the Hamiltonian. This projection takes $\sigma^+ \sigma^- \mapsto \mathbb{I}$ and leaves $\sigma^x$ unchanged such that we are left with an independent boson model. The steady state of $H_{QZE}$ in this case corresponds to a mixture of the eigenstates of $\sigma^x$; any such mixture corresponds to a population of 0.5. 
	 
	 We can also note that the Hamiltonian is symmetric under the unitary transformation that leads to:
	 \begin{align}
	     \{\sigma^x,\sigma^y,a\uq\} &\rightarrow \{- \sigma^x,-\sigma^y,-a\uq\}.
	 \end{align} 
	We would expect the steady state to also obey this symmetry meaning the $\sigma_x$ eigenstates must be equally populated. We can then conclude that in the strong coupling limit we expect the system steady state to correspond to a completely mixed state, i.e. $(\dyad{+}+\dyad{-})/2$. We have confirmed this prediction for the rightmost points of Fig.~\ref{fig:RCcomp}(b) by checking the corresponding system density matrix.

	\section{Conclusions}\label{Sec:conc}
	We have developed a new, exact technique based on tensor networks for finding the dynamics of a system which is coupled to multiple non-commuting baths. Our method can be used to model a broad range of couplings between the system and multiple baths described by arbitrary spectral densities. 
    This has allowed us to probe previously unexplored regimes of simultaneous strong electron-phonon and light-matter coupling, highlighting the importance of proper treatment of non-additive effects in nonequilibrium open quantum systems. In particular, we illustrate this by considering the incoherent population inversion predicted by Maguire~\emph{et al}~\cite{maguire2019environmental}, which emerges in regimes of strong electron-phonon coupling. 
	
	Using multi-bath TEMPO, and backed by the analytic polaron and reaction coordinate approaches, we have highlighted two regimes of light-matter coupling: the first is a regime of \emph{dynamical undressing}, where the light-matter coupling is sufficiently large for optical emission and absorption processes to occur on the same timescales as phonon interactions, preventing the formation of polaron-like states.
	The second regime occurs when the dynamics are dominated by the light-matter interaction, which leads to the optical environment projecting the system into the so-called Zeno subspace, causing decoherence with respect to the basis of $\sigma^x$. This results in steady state populations that are independent of the temperature of the optical field.
	
	The novel behaviour described above is just one example of non-additive physics that emerge in nonequilibrium open quantum systems.
	Multi-bath TEMPO provides a powerful and versatile methodology for studying such phenomena, allowing the simulation of systems that have thus far been intractable to any other techniques.
	Furthermore, the general formulation of TEMPO presented can be applied to fermionic environments, where non-Markovian and non-additive behaviour are expected to play an important role in the dynamical and steady state behaviour of the system of interest~\cite{McConnell2019count,mitchison2018non}.
		
\begin{acknowledgements}
We would like to thank Moritz Cygorek, Jonathan Keeling and Gerald Fux for insightful discussions.
D. G. and D. M. R. acknowledge studentship funding from EPSRC (EP/L015110/1). B. W. L. and E. M. G. acknowledge support from EPSRC (grants EP/T014032/1 and  EP/T01377X/1). A. N.  acknowledges support from an EPSRC Fellowship (EP/N008154/1). J. I.-S. acknowledges
support from the Royal Commission for the Exhibition of
1851.
A. S. acknowledges support from the Australian Research Council Centre of Excellence for Engineered
Quantum Systems (EQUS, CE170100009). 
This work used EPCC's Cirrus HPC Service (https://www.epcc.ed.ac.uk/cirrus). 
\end{acknowledgements}

	\bibliography{paperbib}
	
	\end{document}